\documentclass[12pt]{iopart}

\usepackage[utf8]{inputenc}
\usepackage{iopams}
\usepackage{hyperref}
\usepackage{color}  
\expandafter\let\csname equation*\endcsname\relax
\expandafter\let\csname endequation*\endcsname\relax
\usepackage{mathtools}
\usepackage[]{graphicx}
\usepackage{multicol}

\graphicspath{{./}
	{/home/edi/DATOS/PAPERS_CONFS/P&C_2017/ISHW_2017/Paper/figsPdf/}
}
\DeclareGraphicsExtensions{.pdf, .eps, .png, .jpg, .gif}
\newcommand{\iotab}{\lower3pt\hbox{$\mathchar'26$}\mkern-7mu\iota}

\begin{document}

\title{Oscillatory relaxation of zonal flows in a multi-species stellarator plasma}

\author{E. ~S\'anchez$^1$, I.~Calvo$^1$, J.~L.~Velasco$^1$, F.~Medina$^1$, A.~Alonso$^1$, P.~Monreal$^1$, R.~Kleiber$^2$ {\color{black}and the TJ-II team}}

\address{$^1$Laboratorio Nacional de Fusi\'on / CIEMAT. Avda Complutense 40, 28040, Madrid, Spain.}
\address{$^2$Max-Planck Insitut f\"ur Plasmaphysik, Greifswald, Germany.}
\ead{edi.sanchez@ciemat.es}
\vspace{10pt}

\begin{indented}
\item[]\today
\end{indented}

\begin{abstract}
%Zonal potential perturbations in stellarator plasmas have been shown to have a linear and collisionless oscillatory relaxation with a characteristic oscillation frequency \cite{Mishchenko2008}  smaller than that of the Geodesic Acoustic Mode (GAM). In this work, this
The low frequency oscillatory relaxation of zonal potential perturbations  is studied numerically in the TJ-II stellarator (where it was experimentally detected for the first time). It is studied in full global gyrokinetic simulations of multi-species plasmas. The oscillation frequency obtained is compared with predictions based on single-species simulations using simplified analytical relations. It is  shown that the frequency of this oscillation for a multi-species plasma can be accurately obtained from single-species  calculations using extrapolation formulas. The damping of the oscillation and the influence of the different inter-species collisions is studied in detail. It is concluded that taking into account multiple kinetic ions and electrons with  impurity concentrations realistic for TJ-II plasmas allows to account for the values of frequency and damping rate in zonal flows relaxations observed experimentally.  
\end{abstract}

%\pacs{52.25.Xz, 52.55.Hc, 52.65.−y, 52.55.−s}
\vspace{2pc}
\noindent{\it Keywords}: zonal flows, gyro-kinetic simulations, collisional damping, multi-species
%\submitto{\PPCF}
%\maketitle
\vspace{2pc}

\section{Introduction}
Sheared flows contribute to the regulation of plasma turbulence in magnetic fusion devices via the shear decorrelation effect. Turbulent structures are broken apart as a consequence of a differential rotation thus reducing the anomalous transport produced by them \cite{Biglari90, Hahm88}. 
%The model of turbulence suppression by shear decorrelation was first established for mean sheared radial electric fields \cite{Biglari90} and afterwards extended for oscillating electric fields \cite{Hahm88}. 
Zonal flows (ZFs) are a particular case of sheared flows characterized by zero wavenumber in the plasma potential while they have a finite radial wavelength.
Their capability of  regulating transport an the mechanisms for the drift wave turbulence to produce zonal flows are now well recognized \cite{Diamond2005}. 
% Zonal flows can be generated by turbulence via an inverse cascade of energy from small scales to the large ones thus constituting a self-regulation mechanism for the turbulence. 
However, mechanisms other than turbulence could, in principle, originate zonal flows \cite{Alonso2017}.% that at the end could regulate turbulent transport.

The question of whether ZFs  can 
be sustained in experimental plasmas or are  damped either by collisional or collisionless processes attracted some attention years ago.
Gyro-fluid simulations of turbulence in plasmas predicted high transport levels derived from ITG turbulence  as a consequence of the long term damping of the zonal flows implicit in the gyro fluid models \cite{Glanz1996}.
Rosenbluth and Hinton \cite{Rosenbluth1998a}, using the gyrokinetic formalism, studied for the first time the linear response of a hot (collisionless) plasma 
to a zonal potential perturbation in tokamaks. They treated it  as an initial value problem  and looked for the long term relaxation. They found that 
for times much longer that the  typical bounce time the long-radial-wavelength macroscopic flows are not completely damped, but a residual flow 
survives for long times.
% Simulating the linear relaxation of such a perturbation has became a standard, non-trivial, linear test for gyrokinetic codes.
The work was afterwards extended  to study the collisional case \cite{Hinton1999a} and more realistic tokamak geometries \cite{Xiao2006a},\cite{Xiao2007}.
%Following these seminal works some other authors studied the problem also in axisymmetric geometry \cite{Xiao2006a},\cite{Xiao2007}. 
Sugama and Watanabe \cite{Sugama2005,Sugama2006,Watanabe2008} treated it  
for the case of simplified stellarator configurations finding a link between the neoclassical transport optimization and an enhancement of the residual flows in the so-called inward-shifted configurations of the LHD stellarator. Mynick and Boozer \cite{Mynick2007}, using an action-angle 
formalism in general stellarator geometry, found a similar result.

Mishchenko et al. studied  the collisionless response of a plasma to a potential perturbation in general 
stellarator geometry and found that a  low frequency  oscillation of the zonal potential appears in the relaxation \cite{Mishchenko2008}.
% due to the presence of locally trapped particles with an bounce-averaged radial drift \cite{Mishchenko2008}. 
This work was afterwards extended in \cite{Helander2011a} including the damping of the oscillation.
The linear zonal flow relaxation was studied recently in general stellarator configurations  in \cite{Monreal2016, Monreal2017} and analytical expressions for the residual level and oscillation frequency in stellarators were derived, evaluated numerically and compared to results of gyrokinetic simulations in realistic relevant configurations.
%, finding a very good agreement.

The influence of a radial electric field on the residual level of zonal flows was considered for the first time in \cite{Sugama2009b}, finding that a radial 
electric field has also a beneficial effect because, like the neoclassical optimization,  produces a
reduction of the averaged magnetic drifts of trapped particles.
Global gyrokinetic simulations with the code EUTERPE \cite{Kleiber2010} showed a dependency of the residual 
level  with the electric field in stellarators. 
The influence of the ambient electric field on the low frequency zonal flow oscillation was  studied  analytically and  compared with simulations in \cite{Mishchenko2012b}.

The linear relaxation properties of zonal flows (residual level and oscillation frequency) in stellarators can affect the efficiency of the zonal flows in the regulation of the turbulent transport, which in turn is a non-linear problem. A positive relationship between an increased zonal flow residual and reduced turbulent transport in LHD configurations is found in \cite{WatanabePRL2008, Nunami2012, Nunami2013b}, while in W7-X the residual level appears not to play an important role, but the ZF oscillation frequency seems to be related to the turbulent transport level \cite{Xanthopoulos2011}. These linear properties could provide a way to characterize stellarator configurations in respect of turbulent transport and their evaluation is relatively inexpensive \cite{Monreal2016, Monreal2017}, which makes them appealing to be used in the search for stellarator optimized configurations.

While the residual zonal flow level does not appear as a quantity susceptible of direct measurement \cite{Velasco2012b, Alonso2013, Velasco2013}, the oscillation frequency does, and very recently an oscillatory relaxation of a (zonal) potential perturbation was experimentally detected in pellet injection experiments carried out in the TJ-II stellarator \cite{Alonso2017}. The experimental measurements were compared to global gyrokinetic simulations finding a qualitative agreement between experimental and simulated oscillation frequencies and damping rates. In that work a single-ion plasma was considered in the simulations. In the present work the linear relaxation of zonal potential perturbations, with particular focus on the oscillation frequency and damping rate, is studied in the same magnetic configuration considering collisions and extending the study to multiple-species plasmas.

The rest of the paper is organized as follows. In section \ref{SeccEUTERPE} the code EUTERPE used for the simulations, the equations that it solves and the numerical setup are presented. In section \ref{SecRelaxSinglSpecs} some general properties of the zonal flow relaxation in stellarator geometry are described. In section \ref{SeccZFOscInMultiSpecs} the oscillation frequency of ZFs in a multi-species plasma is studied and simulation results are compared to estimations based on single-species calculations. In section \ref{SeccOscDamping} the damping of the oscillations is studied in some detail and the oscillation frequencies and damping rates are calculated for the experimental conditions of our previous work \cite{Alonso2017}. Finally, section \ref{SecConcl} is devoted to discuss the results and draw some conclusions.
%%%%%%%%%%%%%%%%%%%%%%%%%%%%%%%%%%%%%%%%%%%%%%%%%%%%%%%%%%%%%%%%%%%%%%%%%%%%%%%%%%%%%%%%%%%%%%%
%
\section{The EUTERPE code and the numerical set-up} \label{SeccEUTERPE}
%
%%%%%%%%%%%%%%%%%%%%%%%%%%%%%%%%%%%%%%%%%%%%%%%%%%%%%%%%%%%%%%%%%%%%%%%%%%%%%%%%%%%%%%%%%%%%%%%

EUTERPE is a global gyrokinetic code aimed at simulating plasmas in three dimensional geometries \cite{Jost2001,Kornilov2004a}. It solves the gyroaveraged kinetic equation 
\begin{equation}
\frac{\partial f_a}{\partial t}+ \dot{\mathbf{R}} \frac{\partial f_a}{\partial \mathbf{R}}  + \dot{v_{\|}} \frac{\partial f_a}{\partial v_{\|}} = C(f_a)
\label{KineticEqnWColls}
\end{equation}
for the distribution function $f_a$ of up to three kinetic species. Here $\dot{x} {:=} \frac{\rm d x}{\rm d t}$ means time derivative. 

In the electrostatic approximation here used, the equations of motion for the species $a$ ($\dot{\mathbf{R}}$ and $\dot{{v_{\|}}}$) can be written as $\dot{\mathbf{R}} = \dot{ \mathbf{R}}^0+ \dot{ \mathbf{R}}^1$ and $ \dot{ v}_{\|}  = \dot{ v}_{\|}^0  + \dot{ v}_{\|}^1$, with
\begin{eqnarray}
\label{MotEqtns}
\dot{\mathbf{R}}^0 &=&  v_{\|} \mathbf{b} +  \frac{\mu B  + v_{\|}^{2}} {B^*_{a} \Omega_a} \mathbf{b} \times \nabla \mathbf{B} +
 \nonumber\\
& &\quad\quad\quad\quad\quad	+ \frac{v_{\|}^2  } {B^*_{a} \Omega_a} (\nabla\times \mathbf{B})_{\perp} - \frac{ \nabla  \phi_{LW}  \times \mathbf{b}  \nonumber }{B^*_{a}}\nonumber\\ 
\dot{\mathbf{R}}^1 &=& - \frac{ \nabla  \{\phi\}_G  \times \mathbf{b}  \nonumber }{B^*_{a}}\nonumber\\
\label{MotEqtns2}
\dot{ v_{\|} }^0 &=& - {\mu} \left[ \mathbf{b}
+  \frac{v_{\|}}  {B^*_{a} \Omega_a}(\nabla \times \mathbf{B})_{\perp}
\right] \nabla \mathbf{B} \\
\dot{ v_{\|} }^1 &=& -  \frac{q_{a}} {m_{a}}  \mathbf{b} - \nonumber\\
& & -  \frac{q_{a}} {m_{a}} \left[  \frac{ v_{\|} }  {B^*_{a} \Omega_{a}} \left(\mathbf{b} \times \nabla \mathbf{B} +  (\nabla \times \mathbf{B})_{\perp} \right)
\right]
\nabla \{ \phi\}_G \nonumber .
% \label{MotEqtns3}
% {\dot{\mu }} &=& 0\nonumber.
\end{eqnarray}

The magnetic moment per unit mass, $\mu$, is a constant of motion ($\dot{\mu } = 0$); $q_a$ and $m_a$ are the charge and mass respectively of the species $a$,
$\Omega_a=q_aB/m_a$ is the cyclotron frequency, $\mathbf{b}=\mathbf{ B}/ B$ is the unit vector in the magnetic field ($\mathbf{B}$) direction and  
$B^*_{a}= B + \frac {m_a v_{||}}{q_a} \mathbf{b} \cdot \nabla \times \mathbf{b}$. $\{\phi\}_G$ is the gyro-averaged potential introduced in Ref.~\cite{Hahm88}. 
$-\nabla  \phi_{LW} $ represents the ambient, long wavelength, radial electric field, which is in general determined by neoclassical processes in stellarators \cite{Helander2008}. It is important to include it in the simulation because it can have an influence on the linear relaxation of zonal perturbations \cite{Mishchenko2012b,Kleiber2010}. Here it is included as a zero order contribution. 

In this work, the collisions between different species are taken into account using a pitch-angle scattering collision operator \cite{Kauffmann2010c}. The contributions of inter-species collisions to  $C(f_a) = \sum_bC(f_a, f_b)$ are studied separately in section \ref{SeccOscDamping}.

The code uses a particle-in-cell scheme, the distribution function being discretized using markers whose trajectories 
are given by Equations~\ref{MotEqtns2}. 
The $\delta f$ method is used, so that the distribution function is separated into an equilibrium plus a time-dependent perturbation as $f_a(\mathbf{R}, v_{||}, \mu, t) =  f^{0}_a(\mathbf{R}, v_{||}, \mu)+  \delta f_a(\mathbf{R}, v_{||}, \mu, t).$ A local Maxwellian distribution is used as equilibrium distribution function $f^0_a$.
Using the $\delta f$ decomposition and linearizing, the kinetic equation 	
\begin{eqnarray}
\label{linearizedVlasovEq}
%	\frac{d \delta f}{d t} & = & 
\frac{\partial \delta f_a}{\partial t} + \dot{\mathbf{R}}^0 \frac{\partial \delta f_a}{\partial \mathbf{R}} +  \dot{ v_{\|}}^0 \frac{\partial \delta f_a}{\partial v_{\|}} = \nonumber\\
\quad \quad \quad \quad \quad  = -  \dot{ \mathbf{R}}^1 \frac{\partial f^0_a}{\partial \mathbf{R}} -  \dot{ v_{\|}}^1 \frac{\partial f^0_a}{\partial v_{\|}} + \sum_bC(f_a, f_b)
% -  \frac{d v_{\perp}^1}{d t} \frac{\partial f^0}{\partial v_{\perp}}
\label{linearizedKineticEqn}
\end{eqnarray}
is obtained. 

Two coordinate systems are used in the code: a system of magnetic coordinates (PEST) $(s, \theta^*, \varphi )$ is used for the electrostatic potential and cylindrical coordinates $(r, z,\varphi )$ are used for pushing the particles, where $s=\Psi / \Psi_0$ is the normalized toroidal flux and $\theta^*$ and $\varphi$ are the poloidal and toroidal angles. 
The equation for the electric field is discretized using finite elements (B-splines) and the PETSc library is used for solving it. 
More details about the code can be found in the Refs~\cite{Jost2001,Kornilov2004a,Kornilov2005,Kleiber2006,Kleiber2012a,Borchardt2012}

The simulations are initiated with a radius dependent perturbation to the ion distribution function 
	of the form $\delta f_i \propto w(s) f_M$, where $f_M$ is a local Maxwellian distribution and $w(s)$ is a function dependent of the radius. 
We use two different functions $w(s$) in this work. For the simulations in section \ref{SecRelaxSinglSpecs} to study the influence of the radial scale we use $w(s)= \epsilon cos(k_s s)$, with $k_s$ defining the radial scale of the perturbation and $\epsilon$ being a small quantity ($10^{-3}$ is used here\footnote{As the simulations are linear and we look for the normalized potential this factor does not affect the results.}) to make the $\delta f$ part small as compared to the equilibrium distribution function. In order to emulate the experimental conditions in pellet injection experiments studied in \cite{Alonso2017} we  use a different initial condition, with:
\begin{equation}
%		 F = F_{iM} + \delta f_i; \quad\mbox{ with } 	 \delta f_i = w(r) F_{iM}  \mbox{ and }
		 w(s) = \left \{ 
		 \begin{matrix} \epsilon \quad \quad\quad \quad\quad  \quad\quad \quad  \mbox{in } |s-s_0 | <\delta s \\ 
		  RND(-\epsilon/2, \epsilon/2) \quad \mbox{in } |s-s_0 | > \delta s 
		  \end{matrix}\right. ,
\label{EqInitialCondPellets}
\end{equation}
which represents a density perturbation radially localized around $s_0$. A value of  $10^{-2}$ is used for $\delta s$. 

The linearized equation (\ref{linearizedKineticEqn}) is solved together with the quasi neutrality equation (QNE) that, assuming a long wavelength approximation ($k_{\perp}\rho_i<1$) and adiabatic electrons, reads
% % % % % % % % % % % % % % % % % % % % % % % % % % % % % % % % % % %
\begin{eqnarray}
\label{QNE}
 \frac{e n_{e0} (\phi - \left<{\phi}\right>)}{T_e} - \nabla_{\perp} \cdot \left(\sum_a \frac{m_a n_{a0}}{q_a B^2} \nabla_{\perp} \phi \right) = \nonumber\\
\quad \quad  \quad \quad \quad \quad \quad \quad \quad \quad \quad = \sum_a \{n_a\}_G - \sum_a n_{a0},
\end{eqnarray}
% % % % % % % % % % % % % % % % % % % % % % % % % % % % % % % % % % %
where $\left<{\phi}\right>$ is the potential averaged over a flux surface, $\{n_a\}_G$ is the gyro-averaged density for the ion species $a$ and the $0$ sub-index refers to the equilibrium values. 

The simulations are evolved in time and the zonal potential (m=0, n=0 Fourier component) is monitored during the simulation at several radial positions. Here $m$ and $n$ are the poloidal and toroidal wave numbers, respectively \footnote{The 2D Fourier spectrum (in $\theta*$ and $\varphi$ PEST coordinate angles) of the potential is calculated at each flux surface.}. 
 A typical time evolution of this zonal component is shown in Fig.~\ref{FigZFOscTJIINoCol}.
 
 In order to extract the values of the oscillation frequency and damping rate the potential time evolution is fit to a damped oscillation model like:
\begin{equation}
% \phi_{00}(t)/\phi_{00}(0)=  A \cos({\color{red}\Omega} t {\color{gray}+ \delta}) \exp{(-{\color{blue}\gamma} t)} + R {\color{gray}+ \frac{c}{1+ d t^e}}, 
 \frac{\phi_{00}(t)}{\phi_{00}(0)}=  A \cos(\Omega t + \delta) \exp{(-\gamma t)} + R + \frac{c}{1+ d t^e}, 
 \label{EqFitLFO}
\end{equation}
where $R$ represents the long term residual level, $\Omega$ is the oscillation frequency, $\gamma$ is the damping rate and the last term represents an algebraic decay to the residual level \cite{Helander2011a}. In many practical cases, as those shown in this work, this decay is very fast and the last term in (\ref{EqFitLFO}) can even be neglected in the fit. The fit is carried out with the non-linear fitting function \emph{fit} from the software package Matlab.
 
%%%%%%%%%%%%%%%%%%%%%%%%%%%%%%%%%%%%%%%%%%%%%%%%%%%%%%%%%%%%%%%%%%%%%%%%%%%%%%%%%%%%
%
\section{Zonal flow relaxation in single-species stellarator plasmas} \label{SecRelaxSinglSpecs}
%
%%%%%%%%%%%%%%%%%%%%%%%%%%%%%%%%%%%%%%%%%%%%%%%%%%%%%%%%%%%%%%%%%%%%%%%%%%%%%%%%%%%% 

In this section we use a TJ-II configuration to show some basic characteristics of the linear ZF relaxation in stellarators by means of single-species numerical simulations. 

The linear relaxation of a zonal perturbation to the potential exhibits distinct features in stellarators as compared to the tokamak counterpart. In addition to the GAM oscillation observed in tokamaks \cite{Winsor1968}
a new one, at a smaller frequency, appears in stellarators, which is a completely new phenomenon related with the bounce-averaged radial drift of trapped particles \cite{Mishchenko2008}. We will refer to this feature as the Low Frequency Oscillation (LFO).

Depending on the magnetic configuration one of these oscillations, either the GAM or the LFO, can dominate the relaxation. 
The LFO has a large amplitude in W7-X while the GAM oscillation is almost imperceptible. On the contrary, in LHD (with a much smaller value of $\iotab$ than W7-X) the GAM oscillation is clearly appreciable in simulations, while the LFO is almost undetectable \cite{Helander2011a, Monreal2017}. TJ-II is in between these two cases, and both oscillations are clearly appreciable in simulations, although the damping of the GAM is larger than that of the LFO \cite{Sanchez2013, Monreal2017}. Here we refer to the specific case of equal ion and electron temperatures. As the frequency and damping rates of the oscillations depend also on both temperatures, the situation can be slightly different for different ratios $T_e/T_i$. 

\begin{figure}
	\centering
	%	{\includegraphics[trim=0 0 0 0, clip, height=4cm]{1181-eul_100_44_64_Oscils}}
	%	{\includegraphics[trim=0 0 0 0, clip, height=4cm]{1181-eul_100_44_64_Spectra}}
	{\includegraphics[trim=0 0 0 0, clip, width=8.5cm]{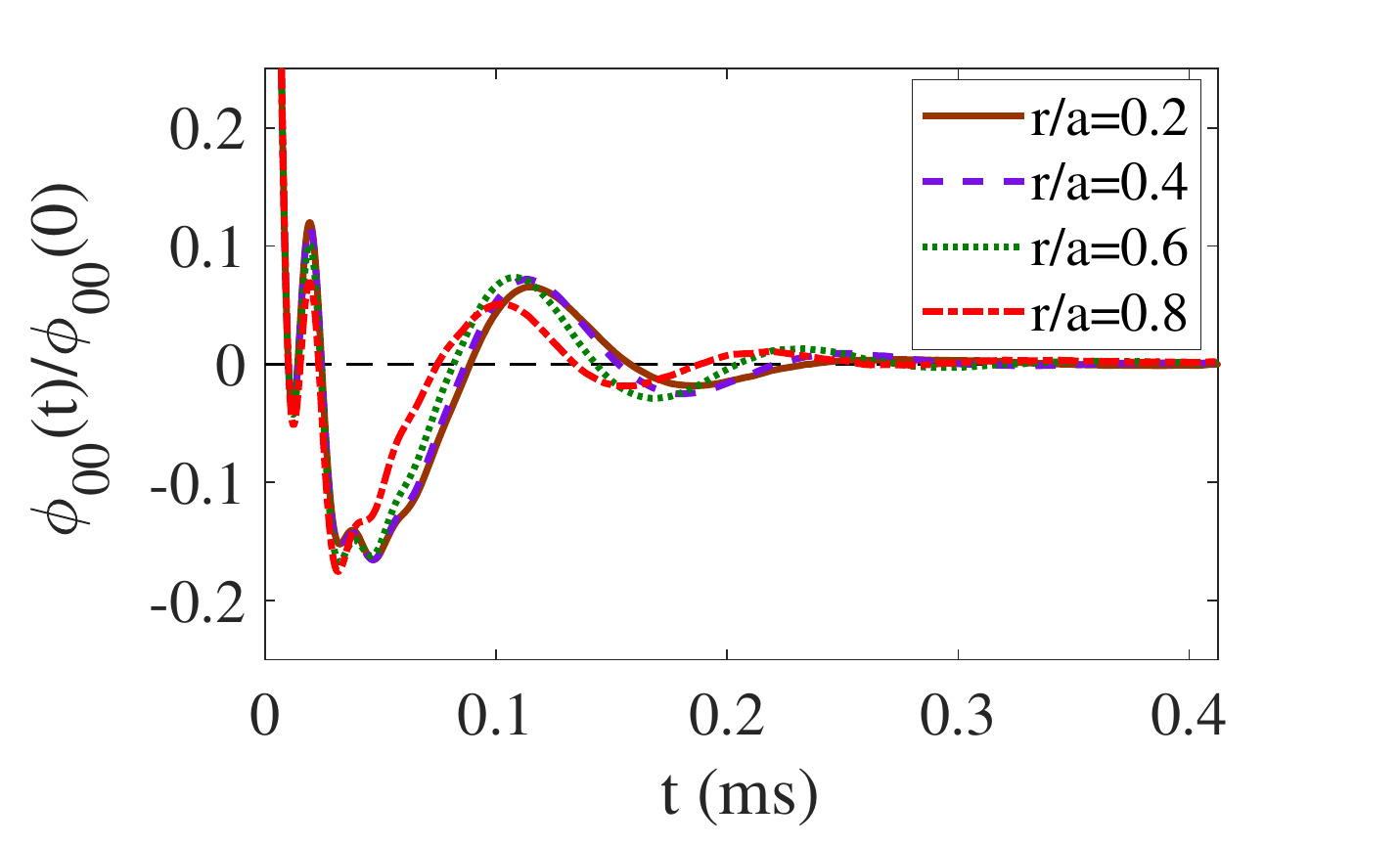}}\\
	{\includegraphics[trim=0 0 0 0, clip, width=8.5cm]{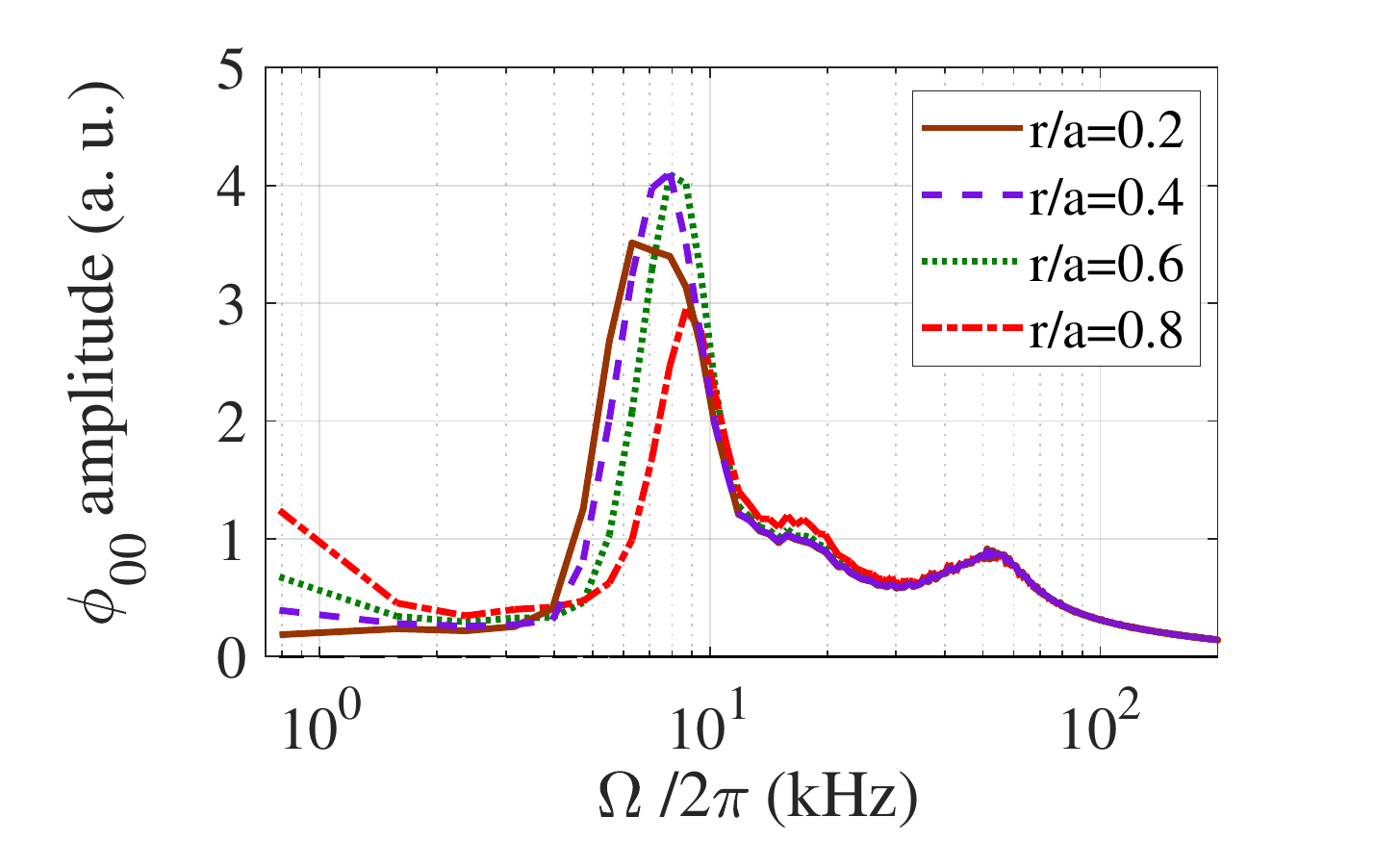}}
	\caption{Time evolution of the zonal potential (normalized to its initial value) in a linear relaxation numerical experiment in TJ-II (top) and Fourier spectra of the potential time traces (bottom). }
	%		The simulation is carried out with $T_e=T_i=100~\rm{eV}$ and $n=10^{19}~\rm{m}^{-3}$.}
	\label{FigZFOscTJIINoCol}
\end{figure}

In Fig.~\ref{FigZFOscTJIINoCol} the time evolution of the zonal potential in a linear relaxation numerical experiment in TJ-II is shown. The simulation is carried out in the standard configuration using flat density and temperature profiles with $T_i=T_e=100~\rm{eV}$ and $n=  10^{19}~\rm{m}^{-3}$. The simulation is initiated with a perturbation to the ion density of the form $w(s)= \epsilon cos(k_s s)$, with $k_s=0.5 \pi$. The zonal  potential  time trace (normalized with its initial value) is shown in Fig.~\ref{FigZFOscTJIINoCol} for several radial positions $r/a= \sqrt{s} = 0.2, 0.4, 0.6, 0.8$. The spectra of these potential time traces, which clearly exhibit two peaks corresponding to the GAM (around 50 kHz) and LFO oscillations (around 6-9 kHz), are also shown in the same figure. As expected for a long wavelength perturbation \cite{Monreal2016} the residual level is close to zero.

%%%%%%%%%%%%%%%%%%%%%%%%%%%%%%%%%%%%%%%%%%%%%%%%%%%%%%%%%%%%%%%%%%%%%%%%%%%%%%%%%%%%
%%%%%%%%%%%%%%%%%%%%%%%%%%%%%%%%%%%%%%%%%%%%%%%%%%%%%%%%%%%%%%%%%%%%%%%%%%%%%%%%%%%%
%
\subsection{Dependency on the radial scale}
%
%%%%%%%%%%%%%%%%%%%%%%%%%%%%%%%%%%%%%%%%%%%%%%%%%%%%%%%%%%%%%%%%%%%%%%%%%%%%%%%%%%%%

The relaxation of the initial perturbation shows a LFO whose frequency reasonably coincides with  semi-analytical calculations \cite{Monreal2017}.  
In order to study the dependency of the oscillation frequency with the radial scale of the perturbation we use a set of simulations carried out in the standard configuration of TJ-II using the same (flat) density and temperature profiles, with $T_i=T_e=100~\rm{eV}$ and $n=  10^{19}~\rm{m}^{-3}$. The simulations are  initiated with a perturbation to the ion density of the form $w(s)= \epsilon cos(k_s s)$, with different radial scales, $k_s=0.5\pi, 1.5\pi, 2.5\pi, 3.5\pi, 4.5\pi$. The frequency is obtained from the fit of the potential time trace to the model of equation (\ref{EqFitLFO}). As we are interested in the LFO and the early times of the time trace are dominated by the GAM oscillation we skip the first part of the time trace in the fit. 

\begin{figure}
	\centering
	{\includegraphics[trim=20 0 40 10, clip, width=8.5cm]{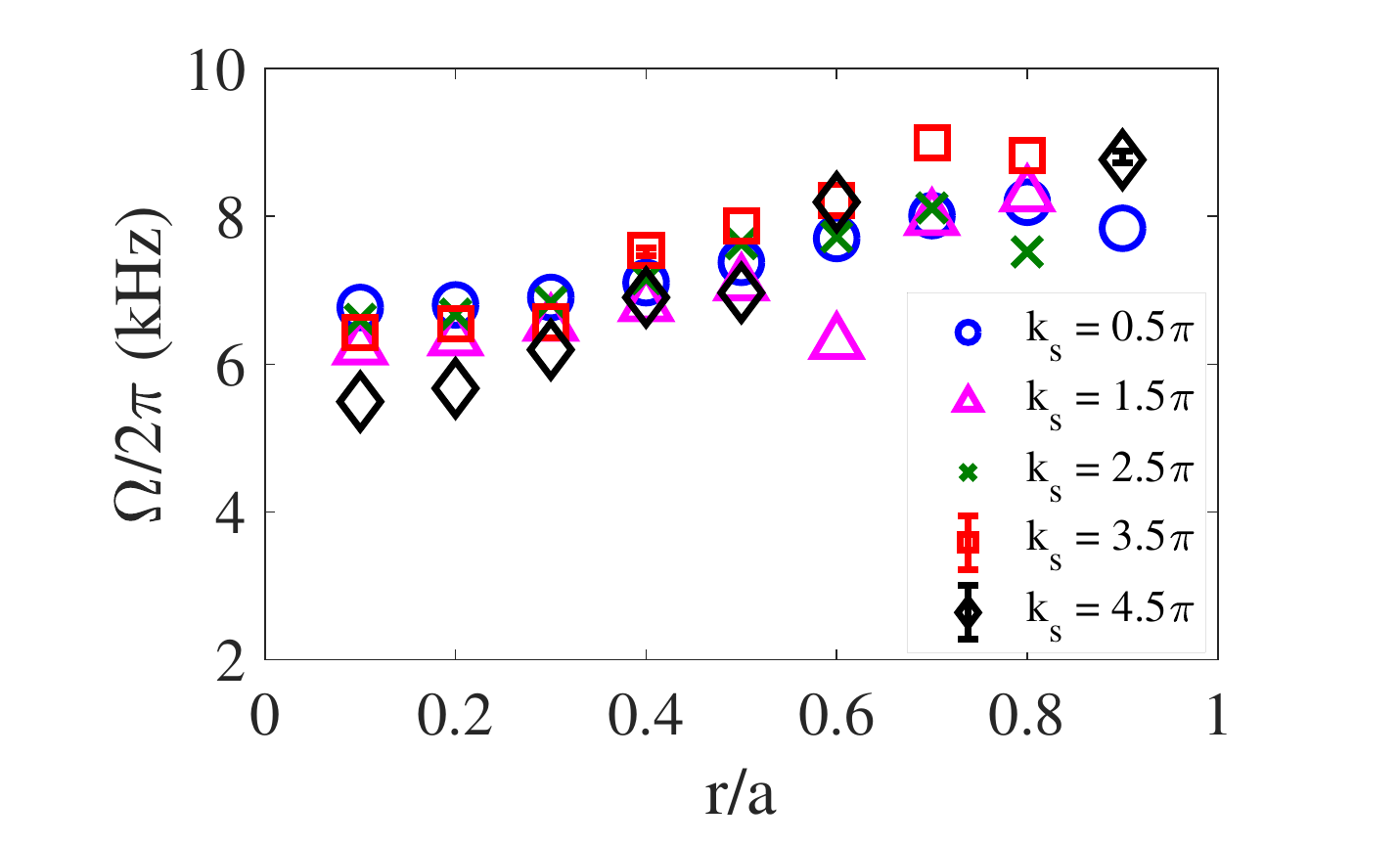}}
	\caption{ZF oscillation frequency vs normalized radius in the TJ-II standard configuration for several radial scales of the initial perturbation, $k_s$.}
%		= 0.5\pi, 1.5\pi, 2.5\pi, 3.5\pi, 4.5\pi$}
	\label{FigZFOscTJIINoColSevsKs}
\end{figure}

In Fig.~\ref{FigZFOscTJIINoColSevsKs}, the oscillation frequency  for this set of simulations is shown versus the radial position. 
In agreement with  the theoretical expectation   \cite{Monreal2017} the frequency is independent of the radial scale of the perturbation in the long wavelength limit, as shown in the figure.
The dispersion in the frequency for different radial scales can be considered within the confidence level of the fitting process. Note that the radial scale of the perturbation varies radially as the radial wave vector is $\mathbf{k_r} = k_s \nabla s$, $|\nabla s|$ having a radial dependency that approximately increases linearly with the radial coordinate $r/a$.
There is a slight increase of the frequency with the radial position that is not related with the radial variation of $k_r $, but with the properties of the magnetic configuration.
%%%%%%%%%%%%%%%%%%%%%%%%%%%%%%%%%%%%%%%%%%%%%%%%%%%%%%%%%%%%%%%%%%%%%%%%%%%%%%%%%%%%
%%%%%%%%%%%%%%%%%%%%%%%%%%%%%%%%%%%%%%%%%%%%%%%%%%%%%%%%%%%%%%%%%%%%%%%%%%%%%%%%%%%%

\subsection{Influence of collisions}

In this section we address the influence of the collisional processes on the oscillation frequency and damping of the  LFO.

\begin{figure}
	\centering
	{\includegraphics[trim=0 0 0 10, clip, width=8.5cm]{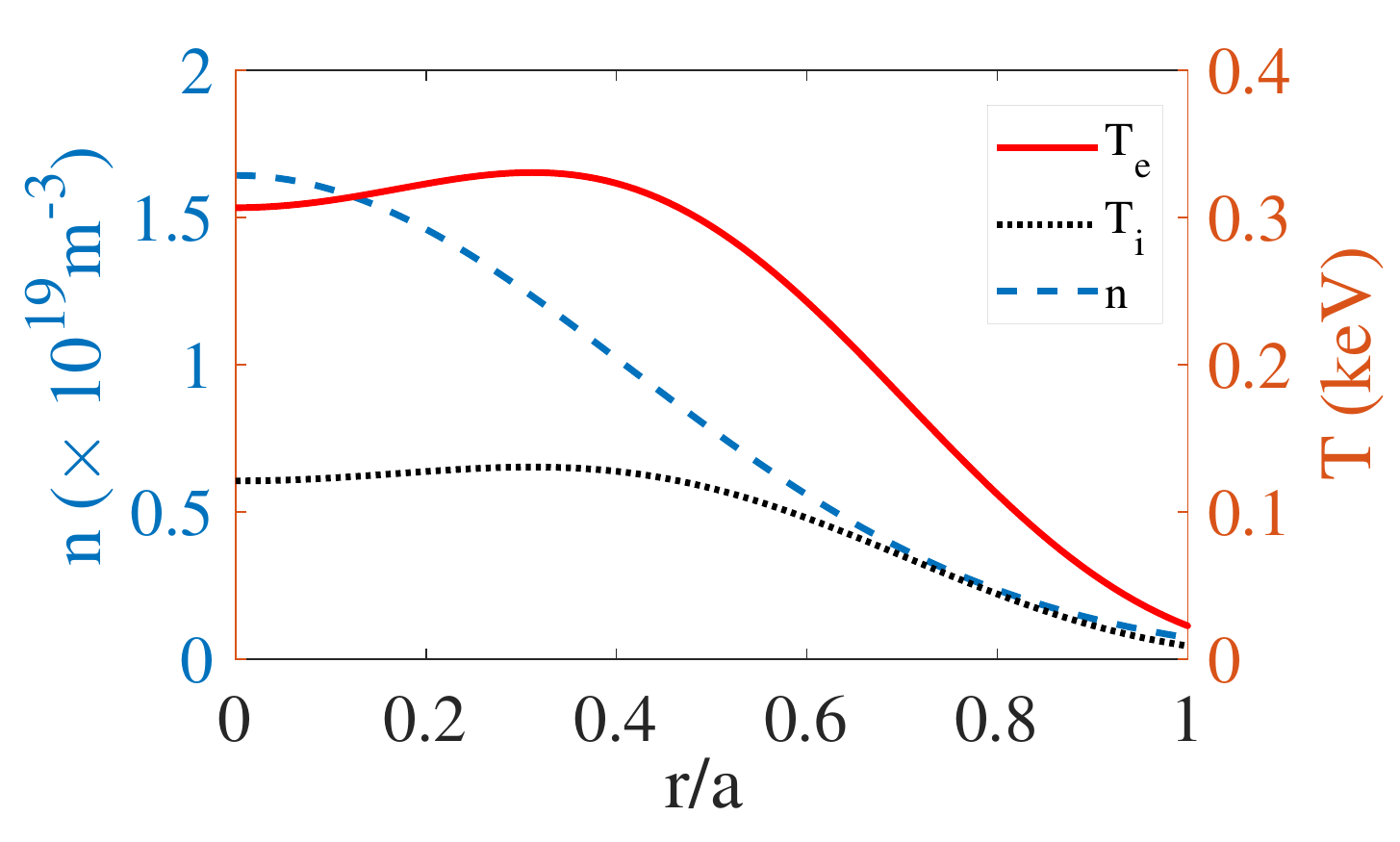}}\\
	{\includegraphics[trim=0 0 0 10, clip, width=8.5cm]{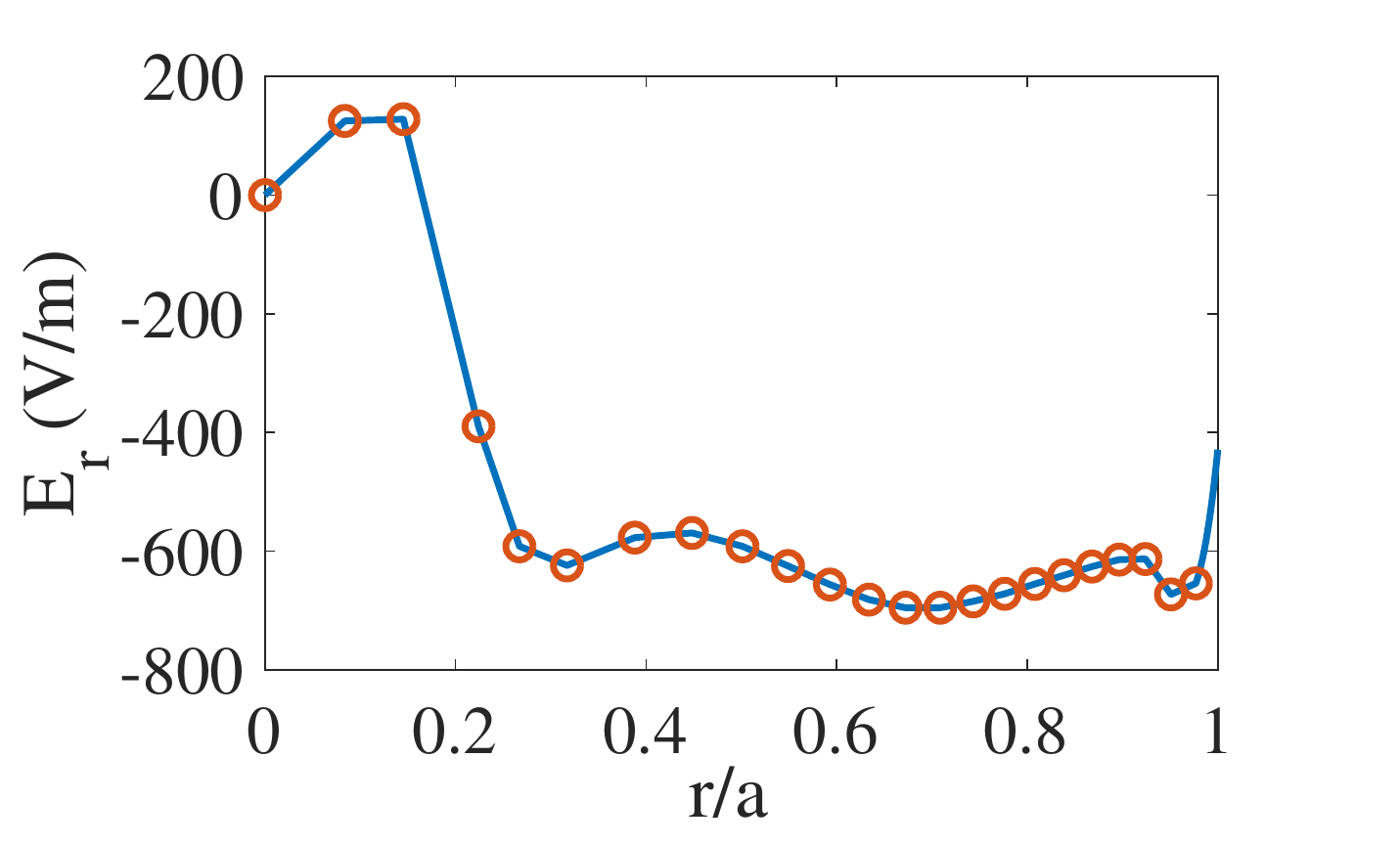}}
	\caption{Density and temperature profiles in the plasma discharge \#39063 of TJ-II (top) and radial electric field ($E_r$) obtained with DKES calculations using these n and T profiles (bottom).}
	\label{FigZ39063NTPofs}
\end{figure}

In order to be close to experimental conditions, in this and following  sections we will use simulations carried out  with experimental profiles from the TJ-II plasma discharge \#39063, corresponding to the experiments of pellet injection \cite{McCarthy2017} analyzed in \cite{Alonso2017}. The density and temperature profiles are shown in Fig.~\ref{FigZ39063NTPofs}. These profiles were reconstructed using an integrated procedure \cite{Milligen2011}, which uses experimental data from the
interferometer, the reflectometer \cite{Estrada2001}, the helium beam \cite{Branas2001},
%, Guzman2009, Tabares2010}
 the Thomson scattering \cite{Herranz2003}
% the CXRS \cite{Carmona2006} 
and from the neutral particle analyzer (NPA) \cite{Fontdecaba2004} diagnostics, when available.
In this way the errors associated to each diagnostic and their calibration factors are taken into account for the global profile reconstruction. It is also shown in the same figure the radial electric field ($E_r$) obtained from  neoclassical calculations with DKES \cite{Hirshman1986,Velasco2012} using those density and temperature experimental profiles. No measurements of the electric field radial profile were available in that series of discharges.
%%%%%%%%%%%%%%%%%%%%%%%%%%%%%%%%%%%%%%%%%%%%%%%%%%%%%%%%%%%%%%%%%%%%%%%%%%%%%%%%%%%%
%%%%%%%%%%%%%%%%%%%%%%%%%%%%%%%%%%%%%%%%%%%%%%%%%%%%%%%%%%%%%%%%%%%%%%%%%%%%%%%%%%%%
 
The simulation in this case is started with the initial condition described in (\ref{EqInitialCondPellets}) to emulate the situation occurring when a pellet is injected in the plasma, which suddenly increases the density at the radial location where the deposition of the particles of the pellet takes place, which is localized radially (see \cite{Velasco2016}). This initial condition in the density produces a perturbation to the plasma potential, which is obtained after the solution of the QNE in the first step of the simulation. The radial profile of this initial zonal perturbation  of the potential (not shown here)
%Note that the potential perturbation  
is long wavelength, which is important to compare with analytical predictions of the oscillation frequency in \cite{Monreal2017}, which are derived under this assumption. Provided that this condition is fulfilled the LFO is a robust phenomenon and its frequency does not depend on the radial scale of the perturbation, as shown in a previous section.

\begin{figure}
	\centering
	{\includegraphics[trim=0 0 10 0, clip, width=8.5cm]{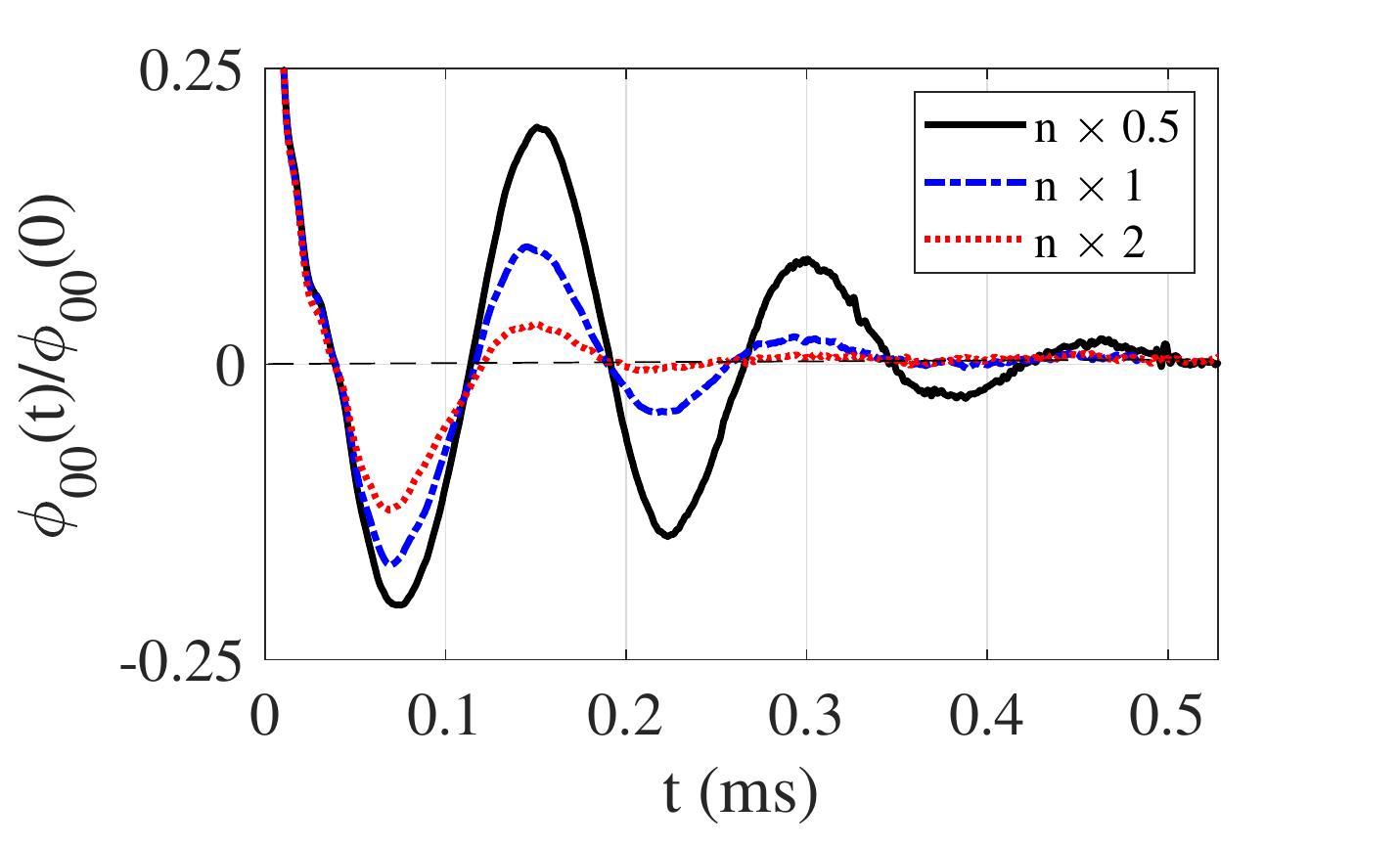}}
	\caption{Normalized time trace of the potential perturbation for three simulations using the profiles of \#39063 in TJ-II with density modified by a factors 0.5, 1 and 2. All cases correspond to the radial position $r/a=0.5$.}
	\label{FigZ39063SevDensitiesOscra05}
\end{figure}

\begin{figure}
	\centering
	{\includegraphics[trim=10 48 10 10, clip, width=8.5cm]{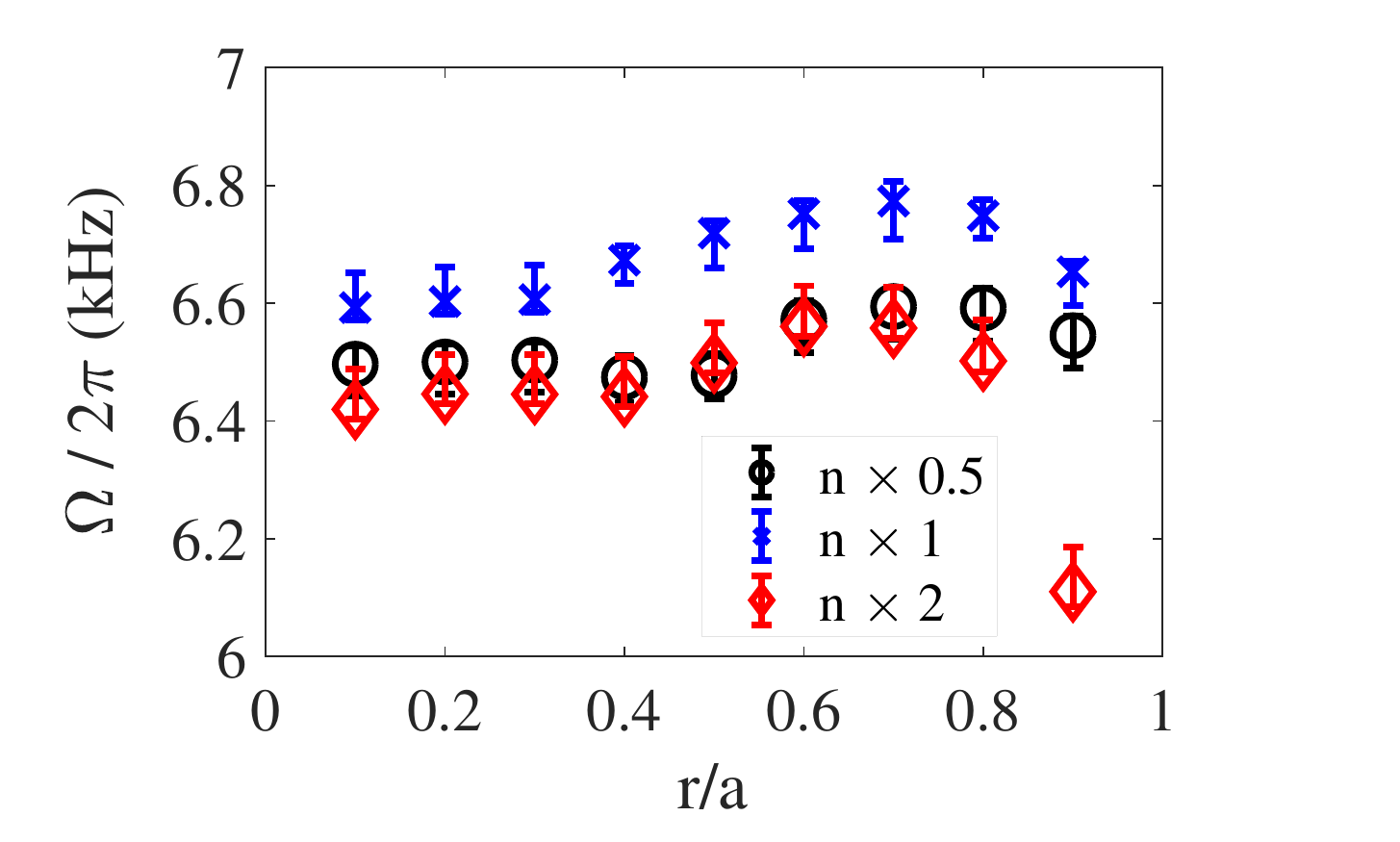}}\\
	{\includegraphics[trim=10 0 10 10, clip, width=8.5cm]{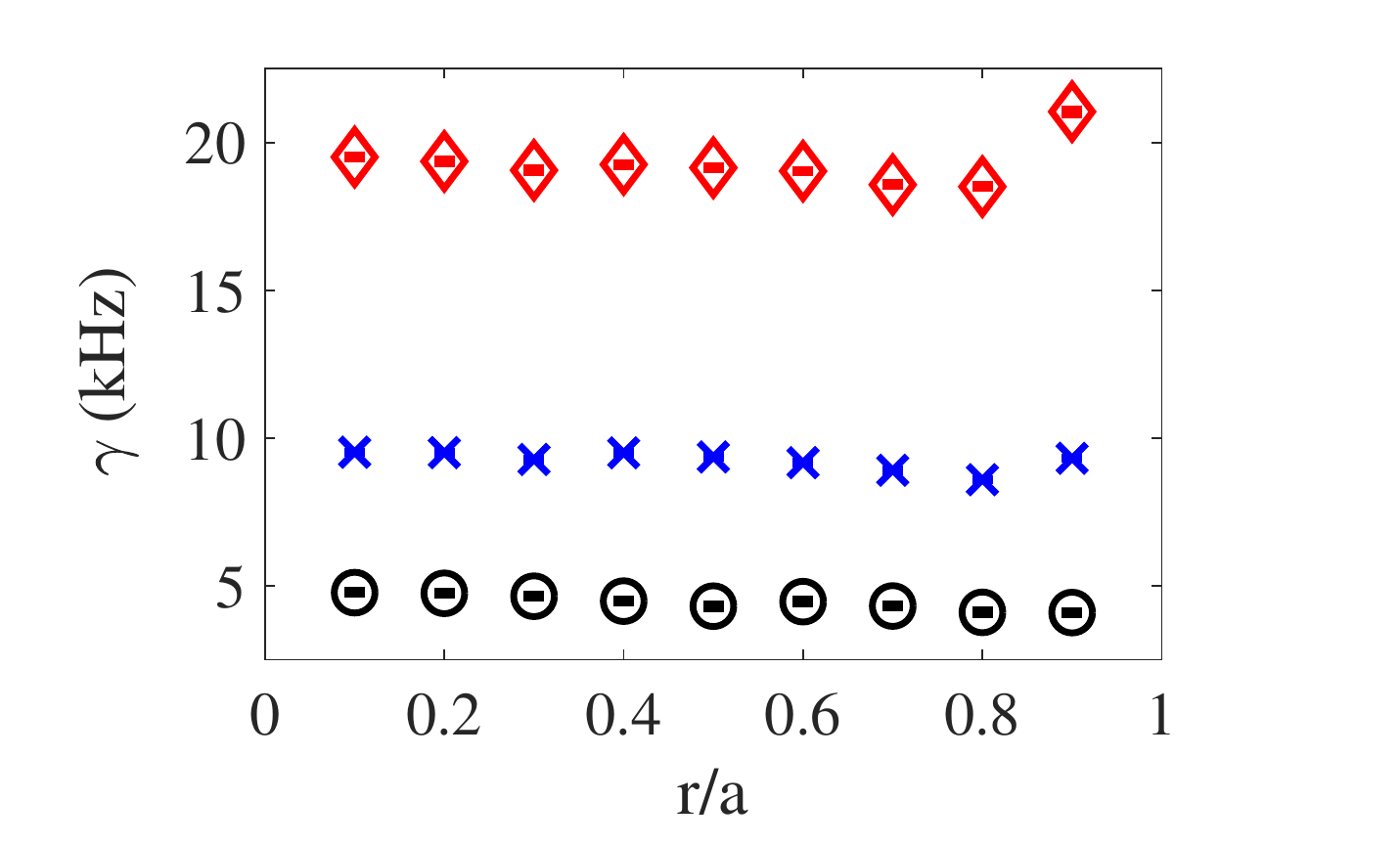}}
	\caption{Frequency (top) and damping rate (bottom) for a set of simulations using the profiles from \#39063 (shown in Fig.~\ref{FigZ39063NTPofs}) with the density modified by several factors: 0.5, 1 and 2.}
	\label{FigZ39063SevDensitiesfgamma}
\end{figure}

We  address the influence of collisional processes on the oscillation frequency by means of a series of simulations in which we change the collision frequency by changing the density while maintaining all the other parameters. We use the  TJ-II standard configuration, the temperature profile shown in Fig \ref{FigZ39063NTPofs} and the density profile in Fig.~\ref{FigZ39063NTPofs} modified by a factor. We carry out three simulations with the actual density in shot \#39063, this density reduced by a factor 2 and finally with it increased by a factor 2. The results are shown in Figs. \ref{FigZ39063SevDensitiesOscra05} and \ref{FigZ39063SevDensitiesfgamma}.
%%%%%%%%%%%%%%%%%%%%%%%%%%%%%%%%%%%%%%%%%%%%%%%%%%%%%%%%%%%%%%%%%%%%%%%%%%%%%%%%%%%%
%%%%%%%%%%%%%%%%%%%%%%%%%%%%%%%%%%%%%%%%%%%%%%%%%%%%%%%%%%%%%%%%%%%%%%%%%%%%%%%%%%%%  
Figure \ref{FigZ39063SevDensitiesOscra05} shows the normalized potential time trace at middle radius, $r/a=0.5$, for the three cases. It is evident from this figure that changing the density does not make a big difference in the oscillation frequency while it strongly affects the damping rate. This is even more clear in figure \ref{FigZ39063SevDensitiesfgamma}, which shows the frequency and damping rate extracted from the fit to the model (\ref{EqFitLFO}) at several radial positions for the three cases. The effect of the collisional processes on the frequency is small, less than 5\%, and there is not a clear tendency with the density (collisionality), while the effect on the damping rate is almost linear with the density.  

%%%%%%%%%%%%%%%%%%%%%%%%%%%%%%%%%%%%%%%%%%%%%%%%%%%%%%%%%%%%%%%%%%%%%%%%%%%%%%%%%%%%
%%%%%%%%%%%%%%%%%%%%%%%%%%%%%%%%%%%%%%%%%%%%%%%%%%%%%%%%%%%%%%%%%%%%%%%%%%%%%%%%%%%% 
It should be taken into account that as the density is increased and the damping rate gets larger the fit is more difficult and its results become less reliable, particularly at the outermost region of the plasma.
In figs. \ref{FigZ39063SevDensitiesOscra05} and \ref{FigZ39063SevDensitiesfgamma} error bars corresponding to the 95\% confidence level in the fit are shown. This error measure, provided by the Matlab fit routine used, is considered to be too small because it does not take into account the constrains imposed to the fitting, which also affect the result, such as the tolerance and the bound limits chosen for the parameters or the time interval at the beginning of the time trace that is skipped from the fit. 
  
\subsection{Influence of the radial electric field}

\begin{figure}
	\centering
	% 	{\includegraphics[trim=0 0 0 0, clip, height=4cm]{CompFreq_ErvsNoEr}}
	% 	{\includegraphics[trim=0 0 0 0, clip, height=4cm]{CompGamma_ErvsNoEr}}
	{\includegraphics[trim=30 0 0 0, clip, width=8.5cm]{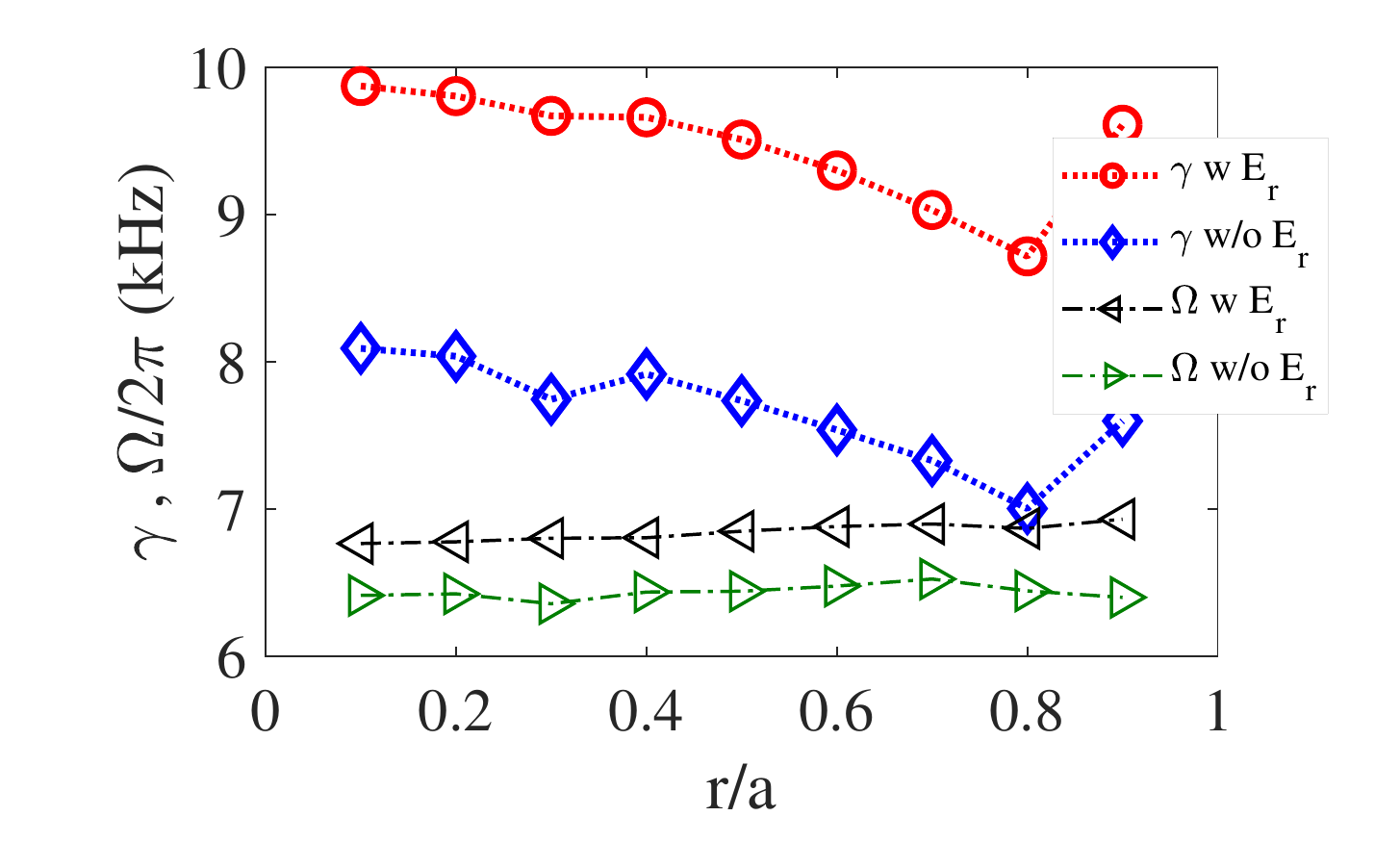}}
	\caption{Frequency and damping rate  for two simulations with profiles from Fig.~\ref{FigZ39063NTPofs} including the NC $E_r$ and without including it. }
	\label{FigZFOscTJIIConySinEr}
\end{figure}

Although  long wavelength ambient electric fields have an influence on the oscillation frequencies, which increase with its strength, for a weak enough electric field there is no dependency of the oscillation frequency on it \cite{Mishchenko2012b}. This is the case in our conditions (Mach number $M<10^{-2}$): including in the simulation the radial electric field shown in Fig.~\ref{FigZ39063NTPofs} does not change much the oscillations frequency, as shown in Fig.~\ref{FigZFOscTJIIConySinEr}.

In this figure we show the damping rate and oscillation frequency in a couple of simulations  using 
the  density and temperature  profiles from shot \#39063 shown before. One of the simulations includes also the neoclassical electric field, while the other one does not include it. 

%%%%%%%%%%%%%%%%%%%%%%%%%%%%%%%%%%%%%%%%%%%%%%%%%%%%%%%%%%%%%%%%%%%%%%%%%%%%%%%%%%%%%
%%%%%%%%%%%%%%%%%%%%%%%%%%%%%%%%%%%%%%%%%%%%%%%%%%%%%%%%%%%%%%%%%%%%%%%%%%%%%%%%%%%%%

From the figure it is clear that for these values of $E_r$ the influence on the oscillation frequency is small (less than 10 \%), while it has a larger influence on the damping rate.

%%%%%%%%%%%%%%%%%%%%%%%%%%%%%%%%%%%%%%%%%%%%%%%%%%%%%%%%%%%%%%%%%%%%%%%%%%%%%%%%%%%%%%%%%%%%
%
\section{ZF oscillation frequency in a multi-species plasma}\label{SeccZFOscInMultiSpecs}
%
%%%%%%%%%%%%%%%%%%%%%%%%%%%%%%%%%%%%%%%%%%%%%%%%%%%%%%%%%%%%%%%%%%%%%%%%%%%%%%%%%%%%%%%%%%%%

In \cite{Monreal2017} an expression for the frequency of the LFO is derived:
\begin{equation}
\Omega=\sqrt{\frac{A_2}{A_1+A_0}},  
\label{EqLowFrequencyDef}
\end{equation}
with
\begin{eqnarray} 
\quad \quad \quad \quad \quad 
A_0 &= &\sum_a {\frac{n_a Z_a^2}{T_a}\left<|\nabla s|^2 \rho_{ta}^2 \right>_r},
\label{EqA0}\\
\quad \quad \quad \quad \quad 
A_1 & = & \sum_a {\frac{ Z_a^2}{T_a}\{\delta _a^2-\overline{\delta _a^2}\}_r }, \label{EqA1}\\
\quad \quad \quad \quad \quad 
A_2 & = & \sum_a {\frac{ Z_a^2}{T_a}{\{\overline{v_{r,a}}}^2\}_r }, 
\label{EqA0A1A2}
%%\left \{ 
%\begin{matrix} A_0 &= &\sum_a {\frac{n_a Z_a^2}{T_a}\left<|\nabla s|^2 \rho_{ta}^2 \right>_r}\\
%%\label{EqA0}
%A_1 & = & \sum_a {\frac{ Z_a^2}{T_a}\{\delta _a^2-\overline{\delta _a^2}\}_r }\quad\\
%%\label{EqA1}
%A_2 & = & \sum_a {\frac{ Z_a^2}{T_a}{\{\overline{v_{r,a}}}^2\}_r } \quad\quad
%\end{matrix} 
%%\right  
%\label{EqA0A1A2}
\end{eqnarray}
where $\rho_{ta}$ is the thermal Larmor radius defined as $\rho_{ta}= v_{ta}/\Omega_a$, with $v_{ta}=\sqrt{T_a/m_a}$ being the thermal velocity of species $a$; $\left< \right>_r $ represents a flux surface average and the over bar  means orbit average. The operation $\{\}_r$ is defined as $\{Q_a\}_r=\left< \int{Q F_{M,a}d^3v} \right>_r$ and $v_{r,a} = v_{d,a} \cdot \nabla s$ is the radial magnetic drift, with
\begin{equation}
v_{d,a} =  \frac{\mu B  + v_{\|}^{2}} {B^*_{a} \Omega_a} \mathbf{b} \times \nabla \mathbf{B} 
+ \frac{v_{\|}^2  } {B^*_{a} \Omega_a} (\nabla\times \mathbf{B})_{\perp},
\label{EqMagnDrift}
\end{equation}
and $\delta_a$ is the radial displacement from the zero-order orbit, which is obtained from the magnetic equation $v_{r,a} =  \overline{v_{r,a}} + v_{\|} \hat {\bf b} \cdot \delta_a$.

From expression (\ref{EqLowFrequencyDef}), which is completely general and valid for an arbitrary number of kinetic species, we can see that all the kinetic species contribute to the oscillation frequency. In the expressions (\ref{EqA0A1A2}) we can separate the purely geometrical contributions from he dependency from species properties as follows:

It is easy to see that $v_{r,a}$ scales with $\frac{v_{ta}^2}{\Omega_a}$ and both  
$\rho_{ta}$ and $\delta_a$ scale with $\frac{v_{ta}}{\Omega_a}$. Then, the quantities 
\begin{eqnarray}
% \left \{
%  \begin{matrix} 
\quad \quad \quad \quad   \xi &=& \frac{\Omega_a^2{\{\overline{v_{r,a}}}^2\}_r}{v_{th}^4}, \quad \quad\quad\quad\quad\quad\\ 
\quad \quad \quad \quad \chi &=& \frac{\Omega_a^2 (\left<|\nabla s|^2 \rho_{ta}^2 \right>_r + \{\delta_a^2-\overline{\delta_a}^2\}_r )}{v_{th}^2}	
%\end{matrix} 
%\right\}
\end{eqnarray}
do not depend on $n_a$, $T_a$ or $m_a $, and we can write ${A_2} \sim   \xi  \sum_a{n_a Z_aT_a}$,  ${A_0 + A_1} \sim \chi \sum_a {n_a Z_a m_a } $, and the oscillation frequency as
\begin{equation}
\Omega \sim  G \sqrt{\frac{\sum_a{n_a Z_a T_a}}{\sum_a{n_a Z_a m_a}}},
\label{OscFreqG}
\end{equation}
with $G = \sqrt{\xi / \chi}$, related only with magnetic geometry.

We can use the expression (\ref{OscFreqG}) to obtain the oscillation frequency for a multi-species plasma based on the frequency for a single-species plasma as: 
\begin{equation}
%\Omega_{iz}(\Omega_{i}) = \sqrt{\frac{(n_iT_i +  \sum_z n_z Z_z T_z)m_i}{(n_im_i + \sum_z n_z Z_z m_z)T_i}}\Omega_{i},
\Omega_{iz} = \sqrt{\frac{(n_iT_i +  \sum_z n_z Z_z T_z)m_i}{(n_im_i + \sum_z n_z Z_z m_z)T_i}}\Omega_{i},
\label{OmegaMultiSfAe}
\end{equation}
where the subindex in $\Omega$ represents the species that is considered as kinetic: $i$ for ions, $e$ for electrons and $z$ for an ion impurity.

As another interesting particular case, the frequency for a plasma with just an ion species and electrons, can be obtained from an adiabatic-electron calculation/simulation as:
%%%%%%%%%%%%%%%%%%%%%%%%%%%%%%%%%%%%%%%%%%%%%%%%%%%%%%%%%%%%%%%%%%%%%%%%%%%%%%%%%%%%
\begin{equation}
%\Omega_{ie}(\Omega_{i}) =  \sqrt{\frac{{(T_i + T_e)m_i}}{{(m_i + m_e)T_i}}} \Omega_{i} \approx  \sqrt{\frac{{T_i + T_e}}{{T_i }}}\Omega_{i},
\Omega_{ie} =  \sqrt{\frac{{(T_i + T_e)m_i}}{{(m_i + m_e)T_i}}} \Omega_{i} \approx  \sqrt{\frac{{T_i + T_e}}{{T_i }}}\Omega_{i},
\label{EqFreqieasfi}
\end{equation}
%%%%%%%%%%%%%%%%%%%%%%%%%%%%%%%%%%%%%%%%%%%%%%%%%%%%%%%%%%%%%%%%%%%%%%%%%%%%%%%%%%%%
 which can be computed at a much cheaper cost, as compared to the case of running a simulation with kinetic electrons.
 
Finally, for the case of a multi-species plasma we can calculate the frequency, including the contribution of kinetic electrons, from the simplest calculation possible: the calculation of the frequency for a single-species with adiabatic electrons, as:
%%%%%%%%%%%%%%%%%%%%%%%%%%%%%%%%%%%%%%%%%%%%%%%%%%%%%%%%%%%%%%%%%%%%%%%%%%%%%%%%%%%%
\begin{equation}
%\Omega_{ize}(\Omega_{i}) = \nonumber\\
%\quad\quad
% = \sqrt{\frac{[n_iT_i +  n_z Z_z T_z + (n_i+ n_zZ_z) T_e]m_i}{[n_im_i +  n_z Z_z m_z + (n_i+ n_zZ_z) m_e]T_i}} \Omega_{i},
%\hspace{-9.1pt}
 \Omega_{ize}\!\! = \!
  \sqrt{\!\!\frac{[n_iT_i +  n_z Z_z T_z + (n_i+ n_zZ_z) T_e]m_i}{[n_im_i +  n_z Z_z m_z + (n_i+ n_zZ_z) m_e]T_i}} \Omega_{i},\!
\label{EqFreqiezasfi}
\end{equation}
%%%%%%%%%%%%%%%%%%%%%%%%%%%%%%%%%%%%%%%%%%%%%%%%%%%%%%%%%%%%%%%%%%%%%%%%%%%%%%%%%%%%
or alternatively, from a calculation for a multi-species plasma with adiabatic electrons, as:
%%%%%%%%%%%%%%%%%%%%%%%%%%%%%%%%%%%%%%%%%%%%%%%%%%%%%%%%%%%%%%%%%%%%%%%%%%%%%%%%%%%%
\begin{equation}
%\Omega_{ize}(\Omega_{iz}) = \nonumber\\
%\quad\quad\quad\quad
%= \sqrt{\frac{n_iT_i +  n_z Z_z T_z + (n_i+ n_zZ_z) T_e}{n_iT_i +  n_z Z_z T_z }}\Omega_{iz}.
\Omega_{ize} =  \sqrt{\frac{n_iT_i +  n_z Z_z T_z + (n_i+ n_zZ_z) T_e}{n_iT_i +  n_z Z_z T_z }}\Omega_{iz}.
\label{EqFreqiezasfiz}
\end{equation}
%%%%%%%%%%%%%%%%%%%%%%%%%%%%%%%%%%%%%%%%%%%%%%%%%%%%%%%%%%%%%%%%%%%%%%%%%%%%%%%%%%%%

In what follows we will validate these shortcuts to obtain the oscillation frequency by comparing the estimation from simple calculations with more complete multi-species simulations, and will take advantage of these properties of the oscillation frequency to extrapolate the frequency for a multi-species real plasma from single-species simulations.

We will use gyrokinetic simulations with EUTERPE in the standard configuration of TJ-II including bulk hydrogen ions an ion impurity to validate  the expression (\ref{OmegaMultiSfAe}). We use the experimental density, temperature and the radial electric field profiles for the discharge \#39063 shown in Fig.~\ref{FigZ39063NTPofs} and include different concentrations of the ion impurity, thus giving effective charge values in the range $1 \leq Z_{eff} \leq 2$, which are in the range of estimations for experimental discharges analyzed in \cite{Alonso2017}, as we will see in \ref{SubSecImpExp}.
 For simplicity, the density profile for the impurity ion, $n_Z$, is assumed equal to that of the bulk ion (hydrogen), $n_0$, scaled by a factor ($n_Z=f n_0$) to match the prescribed effective charge, which can be obtained right from the definition of effective charge as:\\
 % % % % % % % % % % % % % % % % % % % % % % % % % % % % % % % % %
 \begin{equation}
 Z_{eff} = \frac{ n_0 + n_Z  Z^2 }{ n_0 + n_Z Z }  \Rightarrow 
 f = \frac{1 - Z_{eff} } { Z  Z_{eff} - Z^2}.
 \label{EqDensityFactorvsZeff}
 \end{equation}
% % % % % % % % % % % % % % % % % % % % % % % % % % % % % % % % %
%where $n_0$ is the density of the bulk ion.

\begin{figure}
		\centering
	{\includegraphics[trim=20 50 30 10, clip, width=8.5cm]{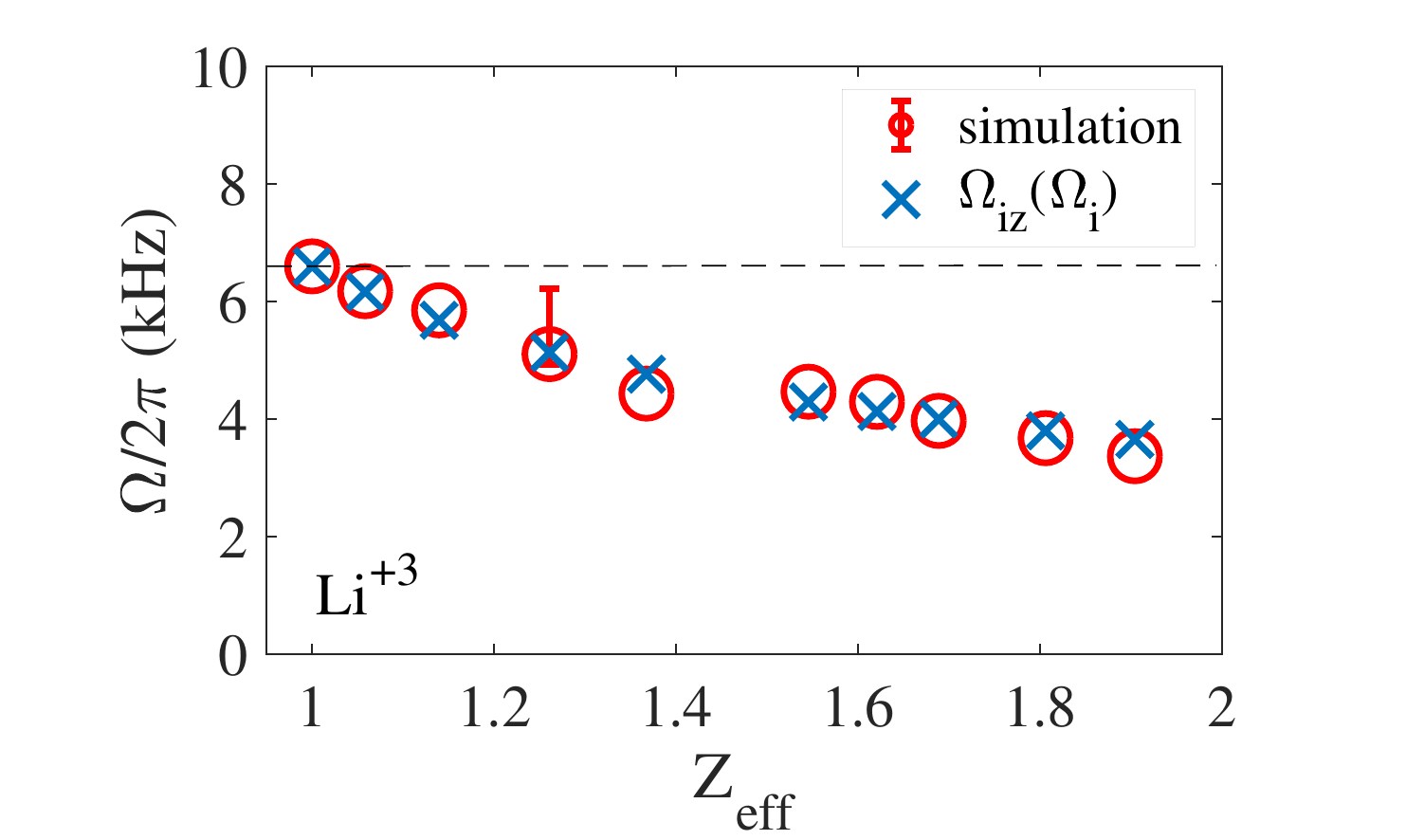}}\\
	{\includegraphics[trim=20 50 30 10, clip, width=8.5cm]{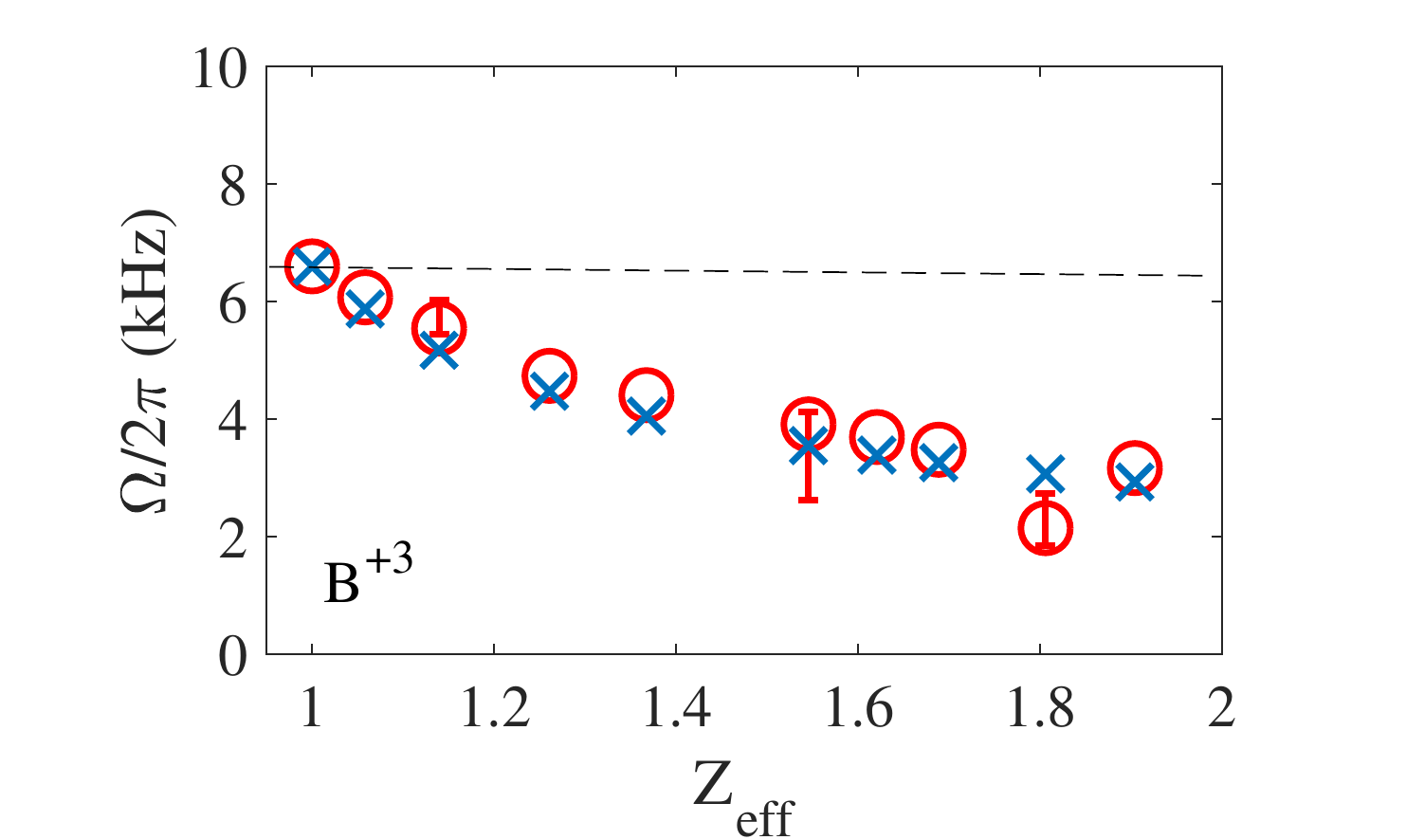}}\\
	{\includegraphics[trim=20 0 30 10, clip, width=8.5cm]{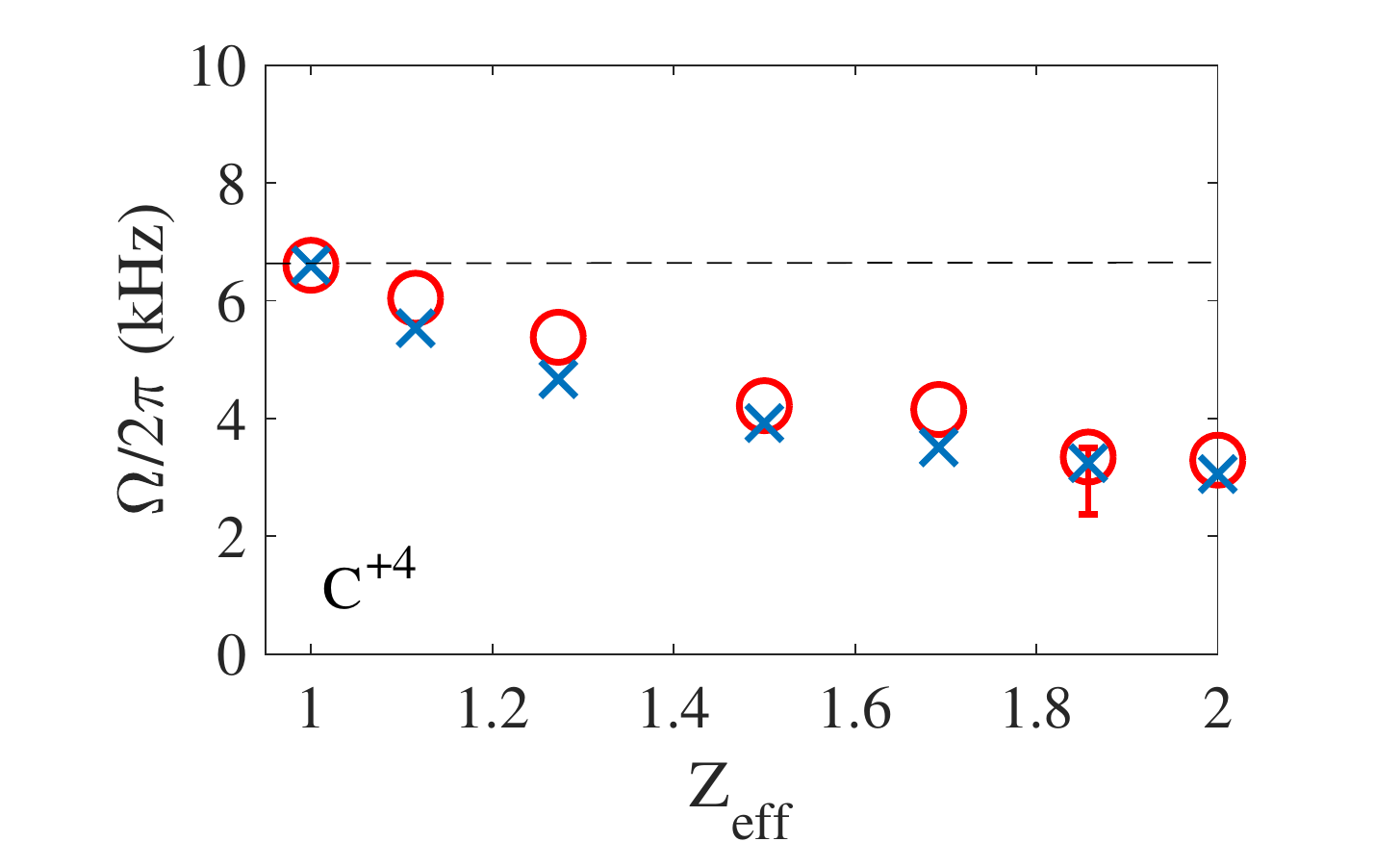}}
	\caption{Oscillation frequency ($\Omega$) vs $Z_{eff}$ for a simulation with two kinetic ion species (hydrogen an a impurity) and adiabatic electrons using the profiles  shown in Fig.~\ref{FigZ39063NTPofs}. Three different kinetic impurities are used: $Li^{+3}$ (top), $B^{+3}$ (center) and $C^{+4}$ (bottom). The circles represent the results of gyrokinetic linear simulations and the plus signs represent the results obtained from expression (\ref{EqFreqiezasfi}) with the $Z_{eff}=1$ case as base for $\Omega_i$. All the data correspond to the radial location r/a=0.5.}
	\label{FigOmegasvsZwExtrap}
\end{figure}

We carry out simulations for three of the most common impurities in TJ-II plasmas: $Li^{+3}$, $B^{+3}$ and $C^{+4}$. The frequency is obtained with a fitting to a damped oscillation model, as explained in Section \ref{SeccEUTERPE}.
The results of the extrapolation using expression (\ref{OmegaMultiSfAe}) and a single species simulation ($Z_{eff}=1$)  are shown in Fig.~\ref{FigOmegasvsZwExtrap} together with the results of the multi-species simulations  ($Z_{eff} > 1$).

%%%%%%%%%%%%%%%%%%%%%%%%%%%%%%%%%%%%%%%%%%%%%%%%%%%%%%%%%%%%%%%%%%%%%%%%%%%%%%%%%%%%
%%%%%%%%%%%%%%%%%%%%%%%%%%%%%%%%%%%%%%%%%%%%%%%%%%%%%%%%%%%%%%%%%%%%%%%%%%%%%%%%%%%%
The points indicated with $\Omega_{iz} (\Omega_{i})$ correspond to those obtained using the expression (\ref{OmegaMultiSfAe}) for different values of $Z_{eff}$. The value $\Omega_i$ here is just that obtained in the simulation with $Z_{eff} = 1$. 

As can be observed in Fig.~\ref{FigOmegasvsZwExtrap} the agreement between the calculation with formula (\ref{OmegaMultiSfAe}) and the multi-species simulations is quite remarkable for all the three cases. The dashed line shows the frequency for the single-species case. 

Once we have singled out the effect of the impurities, we consider now the more realistic case of a multi-species plasma with kinetic electrons, which would be the relevant one for comparison with experimental data.

Running simulations of linear ZF relaxation with kinetic electrons is much more expensive than with adiabatic electrons, because the time step for the integration has to be largely reduced while a long time of simulation has to be run to cover the LFOs. This is even worse in  experimentally relevant cases with electrons hotter than ions and including collisions, which requires an even smaller time integration step as compared to the collisionless case. Collisional kinetic-electron simulations run for such a long time with a short time step resulted numerical unstable in many cases.   For this reason here we only present adiabatic-electron simulations and calculate the frequency for the kinetic-electron case  using the analytic formulas (\ref{EqFreqiezasfi}) and (\ref{EqFreqiezasfiz}) for extrapolation.

\begin{figure}
	\centering
	{\includegraphics[trim=20 47 30 10, clip, width=8.5cm]{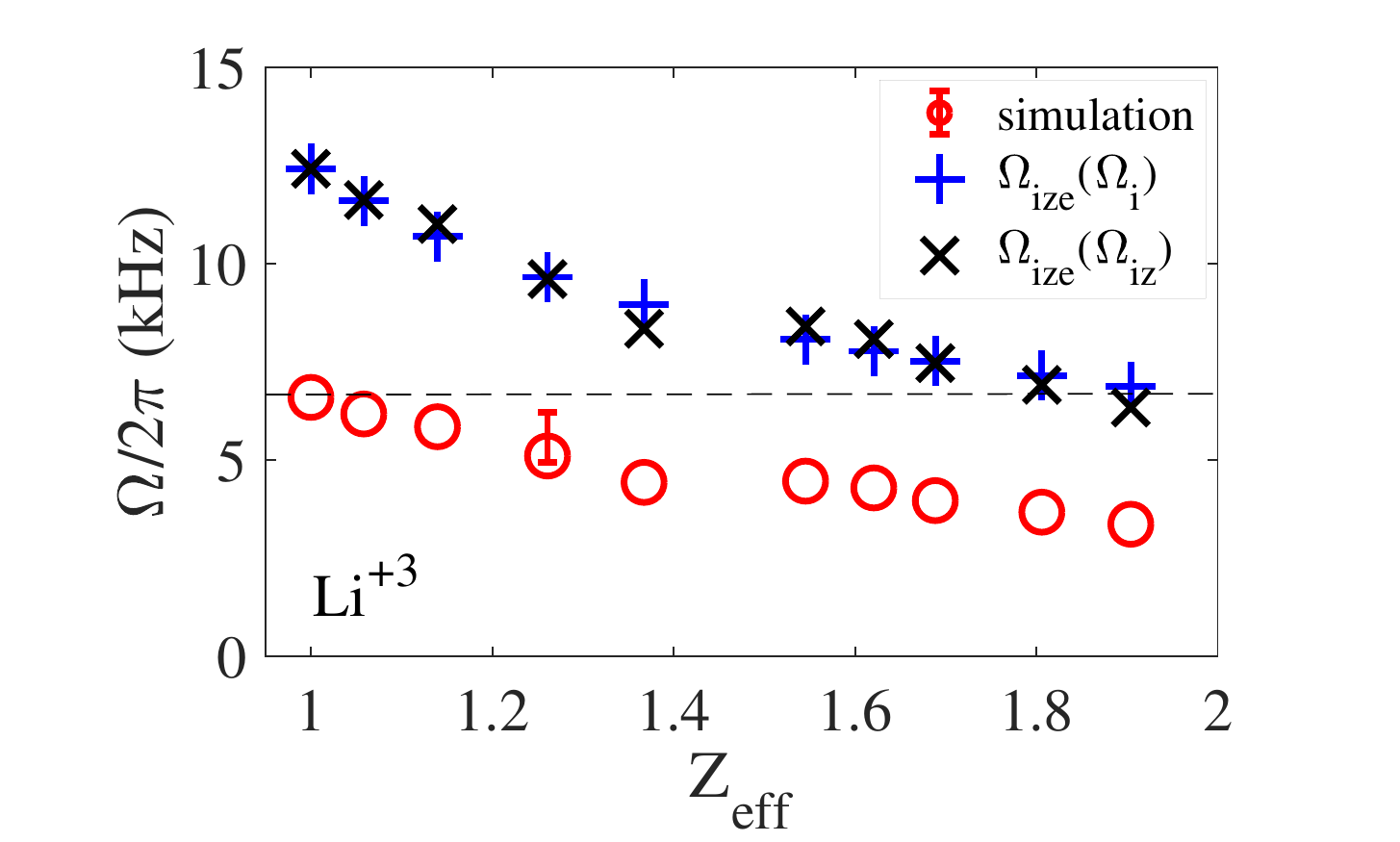}\\
	{\includegraphics[trim=20 47 30 10, clip, width=8.5cm]{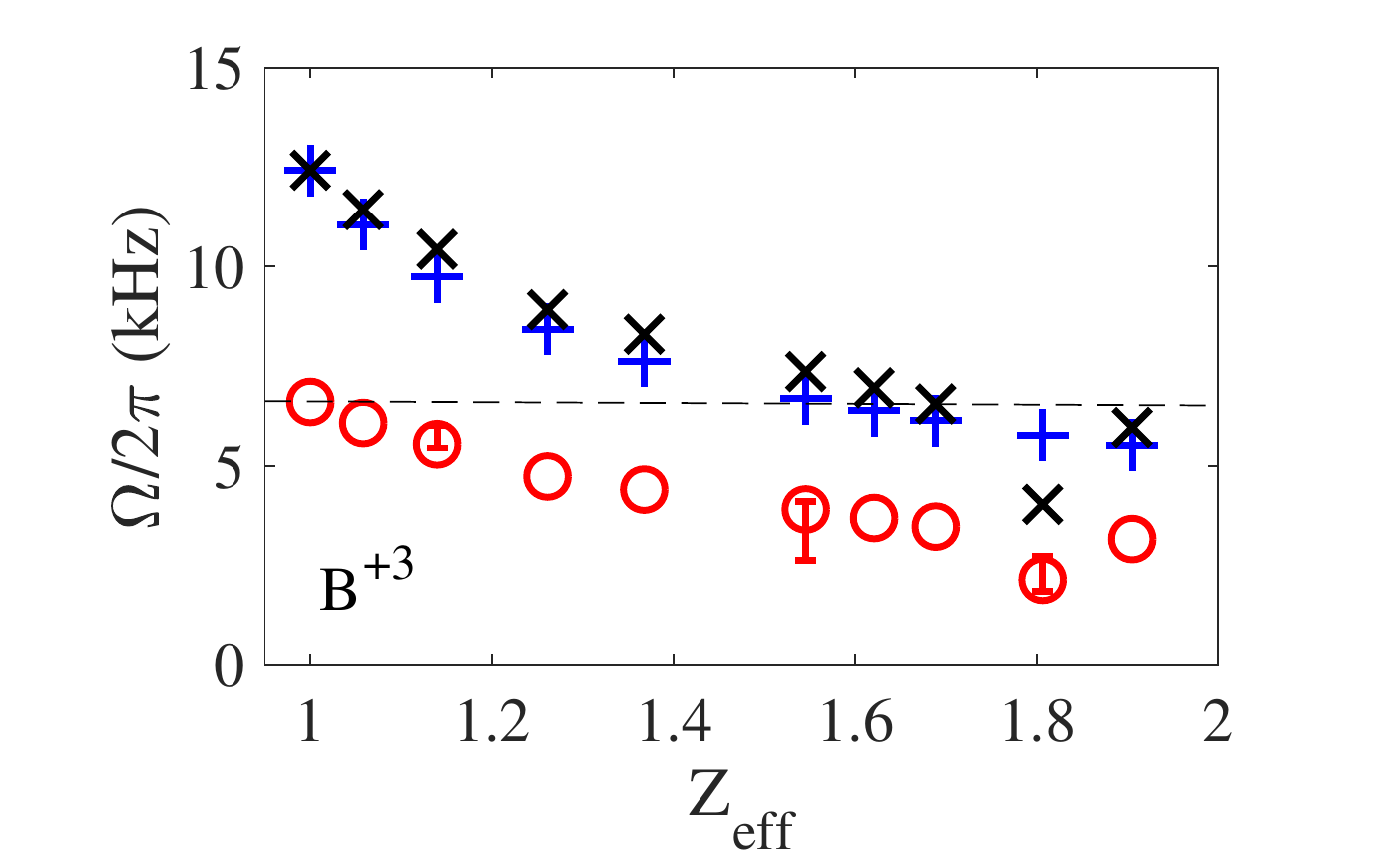}}\\
	{\includegraphics[trim=20 0 30 10, clip, width=8.5cm]{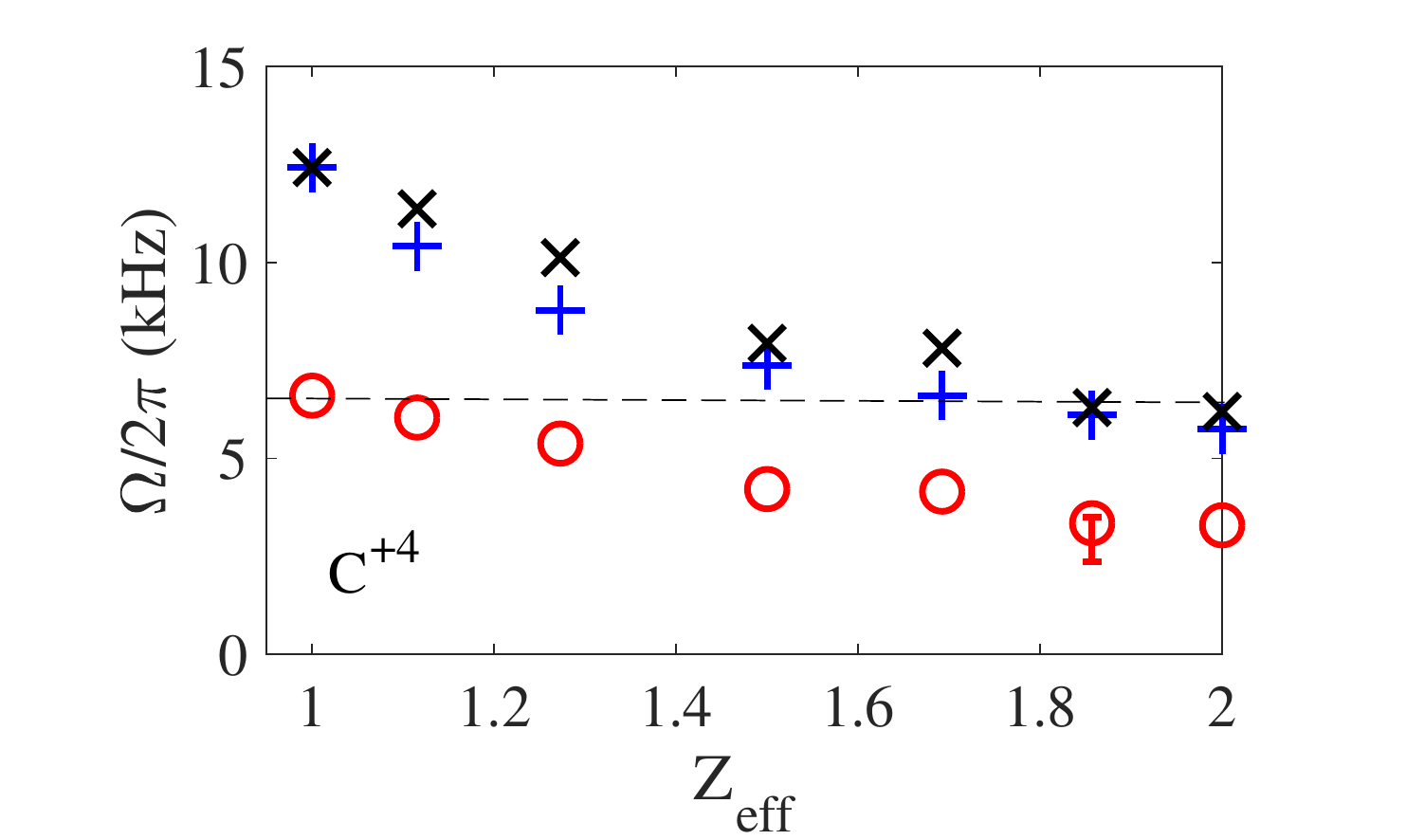}}}
	\caption{$\Omega$ vs $Z_{eff}$ in a multi-species kinetic-electron plasma with the profiles  shown in Fig.~\ref{FigZ39063NTPofs}. The frequency is extrapolated from adiabatic-electron simulations including a kinetic impurity: $Li^{+3}$ (top), $B^{+3}$ (center) and $C^{+4}$ (bottom). The two different extrapolations to the kinetic electron case, using expressions (\ref{EqFreqiezasfi} and (\ref{EqFreqiezasfiz}), are shown. All the data correspond to the radial location r/a=0.5.}
	\label{FigOmegaKEsvsZwExtrap}
\end{figure}

The comparison between estimations of this frequency for a multi-species kinetic-electron plasma using expressions (\ref{EqFreqiezasfi}) and (\ref{EqFreqiezasfiz}) is shown in  Fig.~\ref{FigOmegaKEsvsZwExtrap}  for the same three light species previously used: $Li^{+3}$ (left), $B^{+3}$ (middle) and $C^{+4}$.

%%%%%%%%%%%%%%%%%%%%%%%%%%%%%%%%%%%%%%%%%%%%%%%%%%%%%%%%%%%%%%%%%%%%%%%%%%%%%%%%%%%%
%%%%%%%%%%%%%%%%%%%%%%%%%%%%%%%%%%%%%%%%%%%%%%%%%%%%%%%%%%%%%%%%%%%%%%%%%%%%%%%%%%%%

Again, the dashed line shows the frequency for the single-species adiabatic-electron case. 
It is clear that including electrons increases the frequency with respect to the adiabatic electron case (as can be readily seen from expressions (\ref{EqFreqieasfi}), \ref{EqFreqiezasfi}) and (\ref{EqFreqiezasfiz}), while including impurities reduces it. Both effects compensate partially, the degree of compensation depending on the effective charge. For all the impurity cases, the  oscillation frequency for a real multi-species plasma (with kinetic-electrons) is larger than that obtained by means of single-ion calculations with adiabatic electrons  up to $Z_{eff} \sim 1.6$.

%%%%%%%%%%%%%%%%%%%%%%%%%%%%%%%%%%%%%%%%%%%%%%%%%%%%%%%%%%%%%%%%%%%%%%%%%%%%%%%%%%%%%%%
%
\section{Damping of oscillations in a multi-species plasma}\label{SeccOscDamping}
%
%%%%%%%%%%%%%%%%%%%%%%%%%%%%%%%%%%%%%%%%%%%%%%%%%%%%%%%%%%%%%%%%%%%%%%%%%%%%%%%%%%%%%%%

Here we study the collisional damping of ZF oscillations in the multi-species simulations described in previous section, including the impurities $Li^{+3}$, $B^{+3}$ and $C^{+4}$. 

%%%%%%%%%%%%%%%%%%%%%%%%%%%%%%%%%%%%%%%%%%%%%%%%%%%%%%%%%%%%%%%%%%%%%%%%%%%%%%%%%%%%
\begin{figure}
	\centering
	{\includegraphics[trim=30 47 40 10, clip, width=8.5cm]{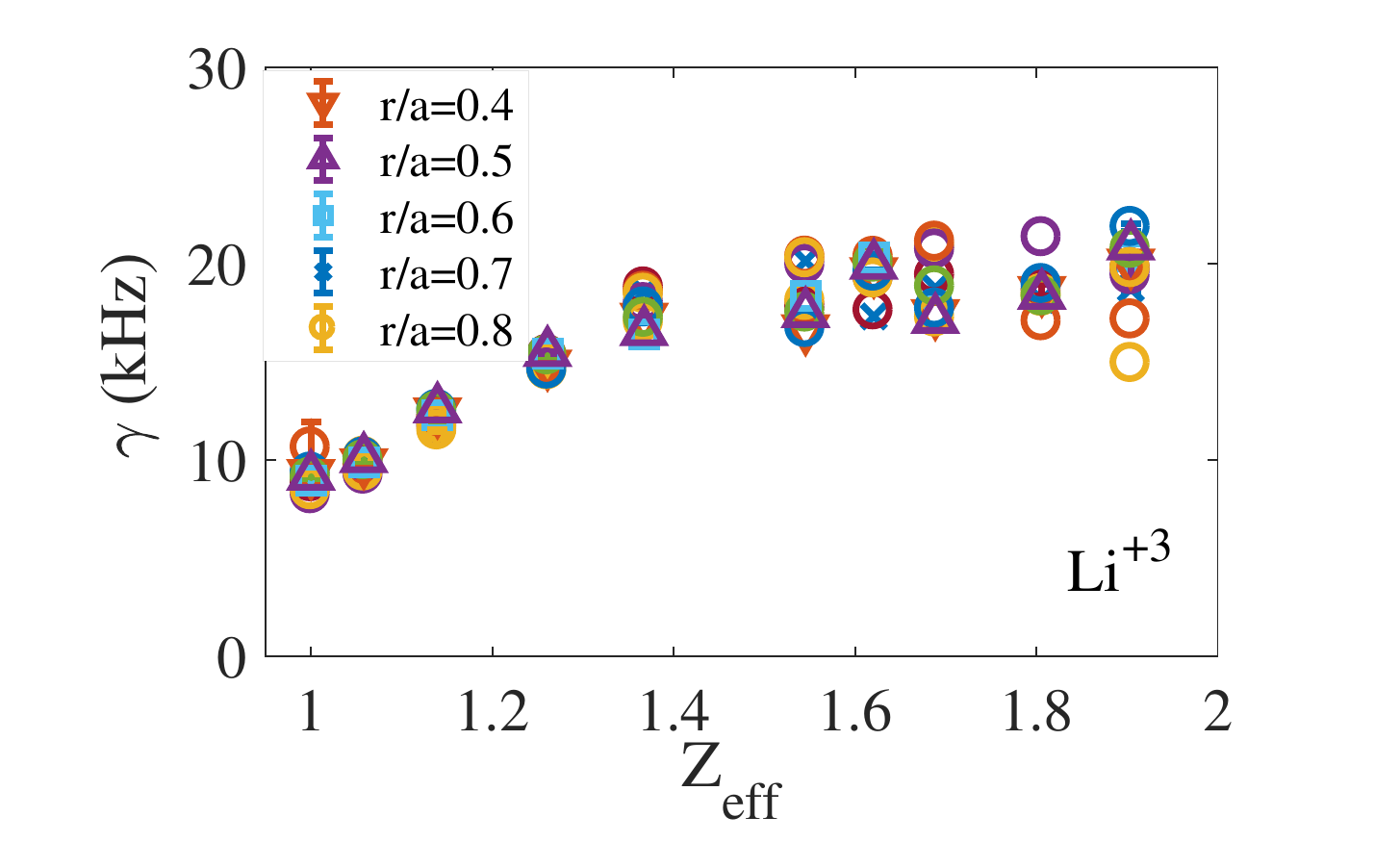}\\
	{\includegraphics[trim=30 47 40 10, clip, width=8.5cm]{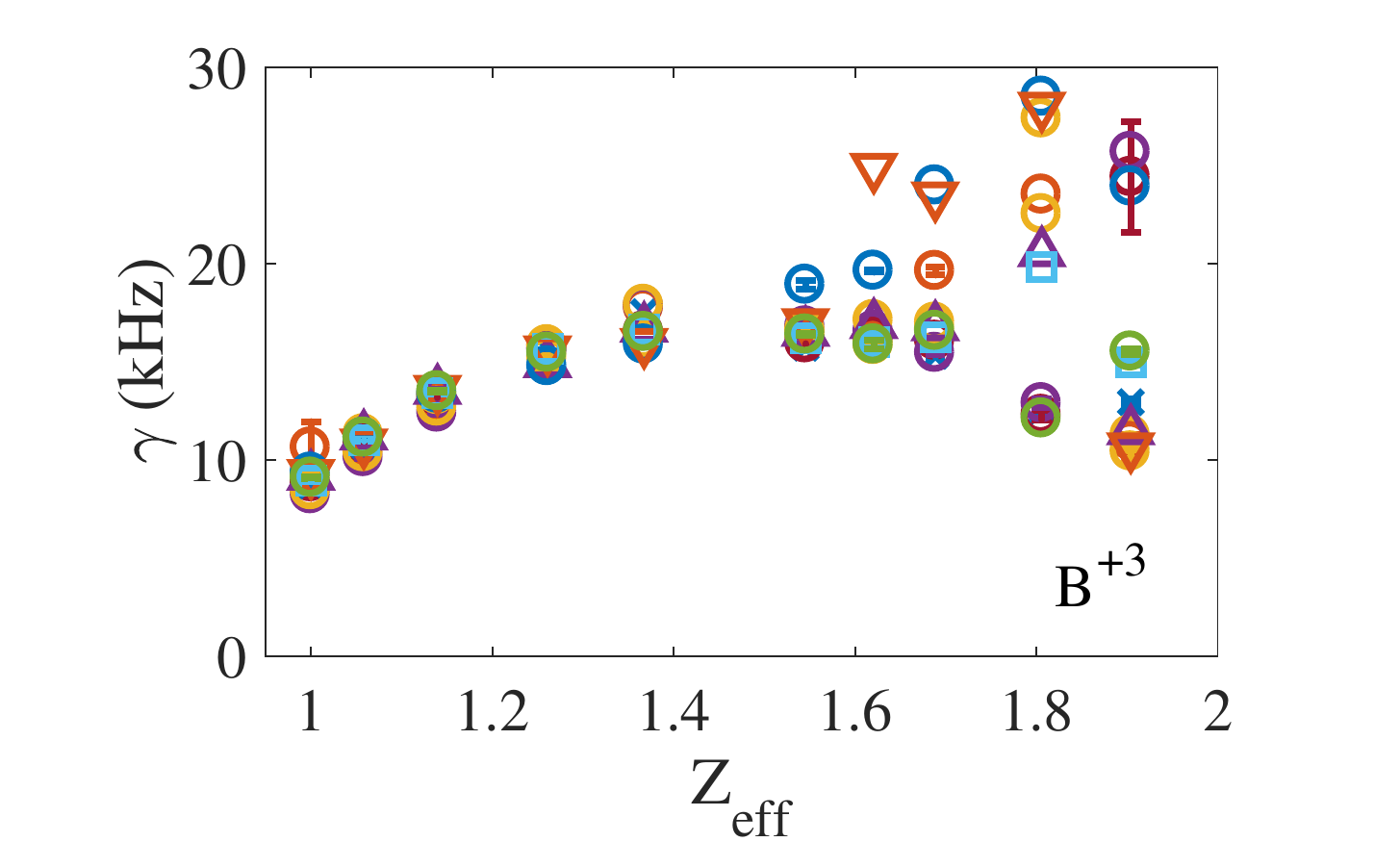}}\\
	{\includegraphics[trim=30 0 40 10, clip, width=8.5cm]{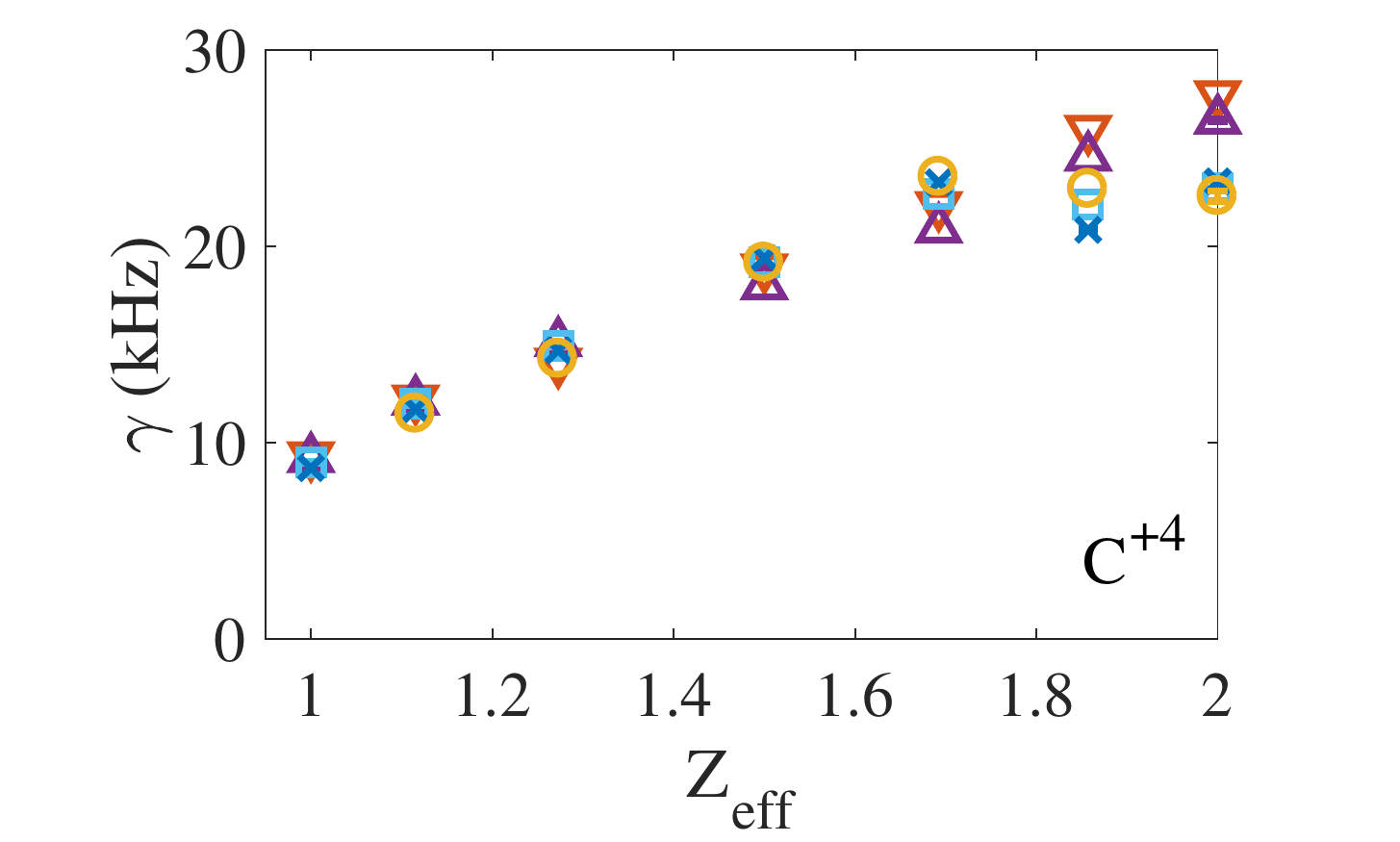}}}   
	\caption{Damping rate ($\gamma$) vs $Z_{eff}$ in multi-species simulations using the profiles  shown in Fig.~\ref{FigZ39063NTPofs} and including a kinetic impurity: $Li^{+3}$ (top), $B^{+3}$ (center) and $C^{+4}$ (bottom).}
	\label{FigDampingFullvsZwExtrap}
\end{figure}
%%%%%%%%%%%%%%%%%%%%%%%%%%%%%%%%%%%%%%%%%%%%%%%%%%%%%%%%%%%%%%%%%%%%%%%%%%%%%%%%%%%%
The damping rate is shown vs the effective charge for several radial locations in Fig.~\ref{FigDampingFullvsZwExtrap}. 

There is a clear increase of the collisional damping rate as the impurity concentration ($Z_{eff}$) increases in all cases. This result is in agreement with those presented in \cite{Braun2009} for the damping of zonal flows in tokamaks, where impurities were found to increase the ZF damping significantly.

 The results for the three impurities studied are very similar. For the highest values of $Z_{eff}$ in Fig.~\ref{FigDampingFullvsZwExtrap} the damping increases notably and the fit of the oscillations to the model (\ref{EqFitLFO}) is less reliable. This is the reason for the dispersion found in the damping rate, particularly in the case of $B$, for values of effective charge $Z_{eff}>1.7$. The error bars provided  by the fitting routine (when available) are smaller than the dispersion of results, which is considered due to the fact that  the fit routine does not take into account all the parameters that affect the fit. As we will see in section \ref{SubSecImpExp} the estimated $Z_{eff}$ for experimental discharges in \cite{Alonso2017} are below 1.8.

\subsection{Contribution of inter-species collisions to the damping}

In this section we study the different contributions to the collisional damping of ZF oscillations separately. We will carry out the study only for the most likely impurity under the conditions of plasmas analyzed in \cite{Alonso2017}, $C^{+4}$. The results shown in the previous section were very similar for the three impurities studied and then similar results in the individual contributions of inter-species collisions can be also  expected for different impurities. The main differences between cases with different impurities in Fig.~\ref{FigDampingFullvsZwExtrap} was for effective charge $Z_{eff}> 1.8$, which are above the experimental values and have a strong damping. %The differences can be  considered to be within the confidence level of the fitting.

\begin{figure}
	\centering
	%	{\includegraphics[trim=30 0 50 0, clip, width=5.25cm]{Pellets_39063_CompTermss025_Gamma-f00Li.pdf}}
	%	{\includegraphics[trim=30 0 50 0, clip, width=5.25cm]{Pellets_39063_InflImpCNoColie_s025_Gamma-f00}}
	{\includegraphics[trim=10 0 15 10, clip, width=8.5cm]{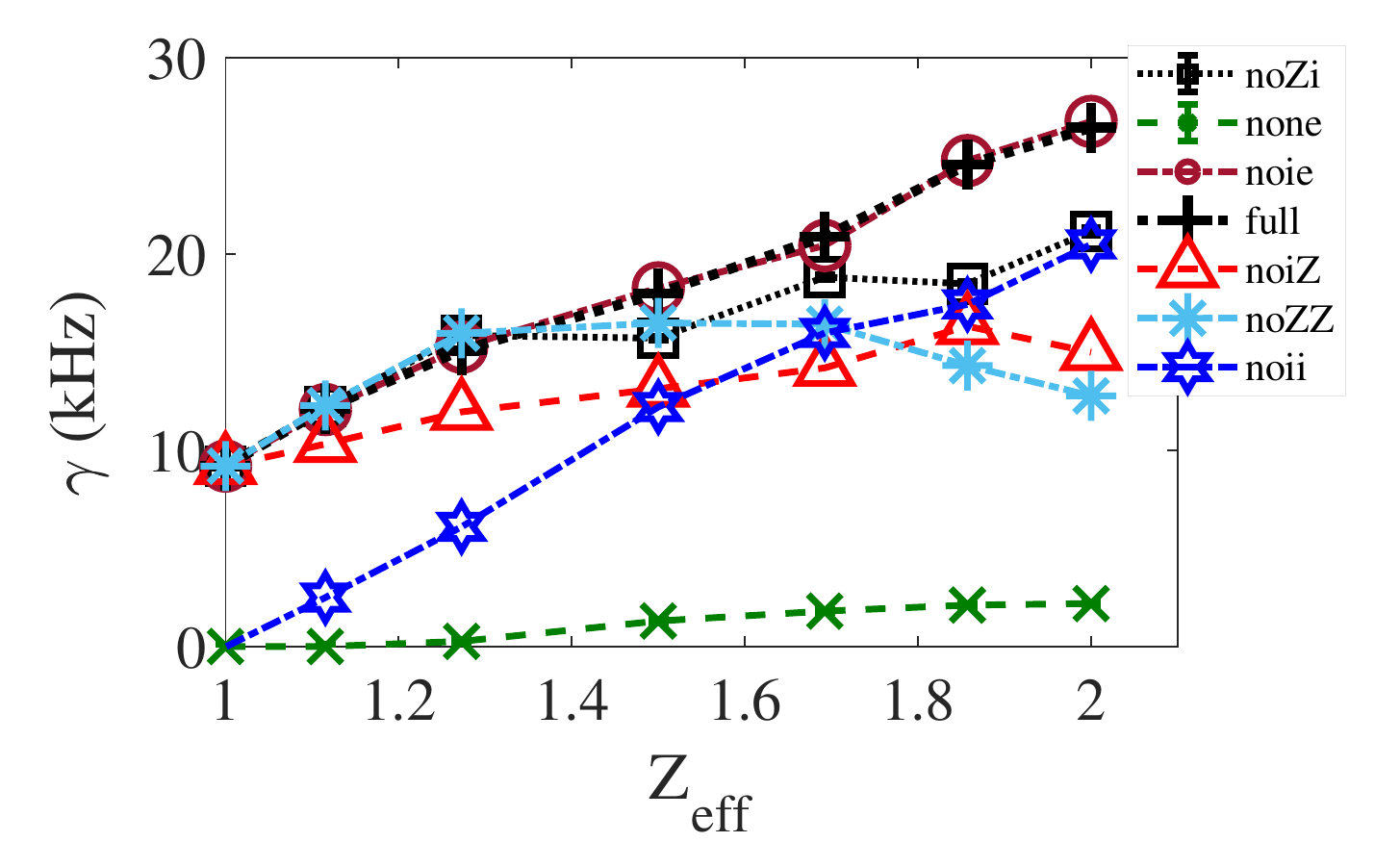}}
	\caption{Damping rate ($\gamma$) vs $Z_{eff}$ for simulations with the profiles  shown in Fig.~\ref{FigZ39063NTPofs} and including $H^{+}$ (bulk) and $C^{+4}$ kinetic ions and  all collisions between bulk ions, electrons and impurity ions ("full"), without any collisions ("none"), excluding $H-H$ collisions ("noii"), excluding $H-C$ collisions ("noiZ"), excluding $C-H$ collisions ("noZi"), excluding $H-e$ collisions ("noie") and  excluding  $C-C$ collisions ("noZZ").}
	\label{FigDampingCompTerms}
\end{figure}

We use the same density, temperature and electric field profiles corresponding to the discharge \#39063, as in previous sections. The damping rate, $\gamma$, is shown versus effective charge, $Z_{eff}$, in Fig.~\ref{FigDampingCompTerms} for several cases in which different inter-species collisions are suppressed. First, it is shown the collisionless case (indicated as "none" in figure) and the full collisional case ("full" in the figure). We show in the same figure five more cases: excluding bulk ion (H-H) collisions ("noii"), excluding C-C collisions ("noZZ"), excluding the collisions between the bulk ion and $C$ ("noiZ"), excluding collisions between $C$ and the bulk ions ("noZi") and finally excluding collisions between the bulk ion and electrons ("noie").
In all cases the same pitch-angle scattering collision operator is used \cite{Kauffmann2010c}.
The results are shown in Fig.~\ref{FigDampingCompTerms} for the specific radial position $r/a=0.5$. There is not an important dependency with radial position, as could be expected from figure \ref{FigDampingFullvsZwExtrap} in which the full damping rates for several radial positions were shown together.

%%%%%%%%%%%%%%%%%%%%%%%%%%%%%%%%%%%%%%%%%%%%%%%%%%%%%%%%%%%%%%%%%%%%%%%%%%%%%%%%%%%%
%%%%%%%%%%%%%%%%%%%%%%%%%%%%%%%%%%%%%%%%%%%%%%%%%%%%%%%%%%%%%%%%%%%%%%%%%%%%%%%%%%%%

It is clear from the figure that the collisional damping is much larger than the collisionless one, by more than a factor 10. The collisional damping is dominated by collisions between bulk ions while those of bulk ions with electrons are almost not relevant, as expected.
The next contribution is that coming from collisions between bulk ions and $C$ for $Z_{eff}<1.7$, while the importance of the  contribution from  $C-C$ collisions increases for $Z_{eff}>1.5$. The contribution of impurities with bulk ions is always very small.

\subsection{Oscillatory relaxation under experiment-relevant conditions}\label{SubSecImpExp}

Finally, in this section we calculate the oscillation frequency and damping rate for multi-species plasmas, including bulk hydrogen ions, a light impurity and adiabatic electrons, under experimentally relevant conditions.  We only present simulations for the discharges that showed clear oscillatory relaxations, which are listed in table \ref{TabZeff}.
 
 In this case, we use the values of effective charge estimated for these discharges from experimental soft X-ray (SXR) measurements. To this end information about the plasma composition is required, which is obtained from four detectors  equipped with four different Be filters \cite{Baiao2010} that respond differently with the plasma composition. Filter thicknesses were chosen to discern the presence of some of the main impurities in TJ-II plasmas (Li, B, C, O and in much less quantity F). The IONEQ code \cite{Weller1987} is used to estimate the SXR emissivities. In this way, $Z_{eff}$ is obtained from the absolute and relative values of the experimental signals, as well as a rough estimation of plasma composition. 
 
 The dominant impurity ion in these plasmas was estimated to be $C^{+4}$. The estimated values of $Z_{eff}$ for the experimental discharges reported in \cite{Alonso2017} as showing an oscillatory relaxation,  always in the range $1 < Z_{eff} < 1.75$, are shown in table \ref{TabZeff}.  These values are central $Z_{eff}$ estimations, which can be considered an upper limit for the effective charge, because the impurity concentration is maximum at the center. 

In the simulations, the experimental density and temperature profiles and neoclassical electric field obtained from them (not shown here) corresponding to the set of plasma discharges studied in \cite{Alonso2017}, \#39047-39064, are used. For simplicity, the density profile for the impurity ion is assumed equal to that of the bulk ion (hydrogen) scaled by a factor to match the prescribed (experimental) effective charge in all radii. The required factor  is obtained using the expression (\ref{EqDensityFactorvsZeff}). We run simulations including $C^{+4}$ as impurity and use adiabatic electrons.
% % % % % % % % % % % % % % % % % % % % % % % % % % % % % % % % %
% % % % % % % % % % % % % % % % % % % % % % % % % % % % % % % % %
\begin{table}\setlength\tabcolsep{2.2pt} 
	\centering
	\caption{Effective charge estimated from experimental SXR measurements in plasma discharges studied in  \cite{Alonso2017}}
	\begin{tabular}{|l|c|c|c|c|c|c|c|c|c|c|c|c|}
		\hline	
		shot &	39047   & 39048 & 39050 &39056 & 39058 & 39063 & 39064\\
		\hline	
		$Z_{eff}$ &1.29 &1.36  &1.30   &1.72 & 1.78    & 1.28  &1.29\\		\hline	
	
	\end{tabular}		\label{TabZeff}
\end{table}
% % % % % % % % % % % % % % % % % % % % % % % % % % % % % % % % %

\begin{figure}
	\centering
	{\includegraphics[trim=55 20 70 20, clip, width=7.cm]{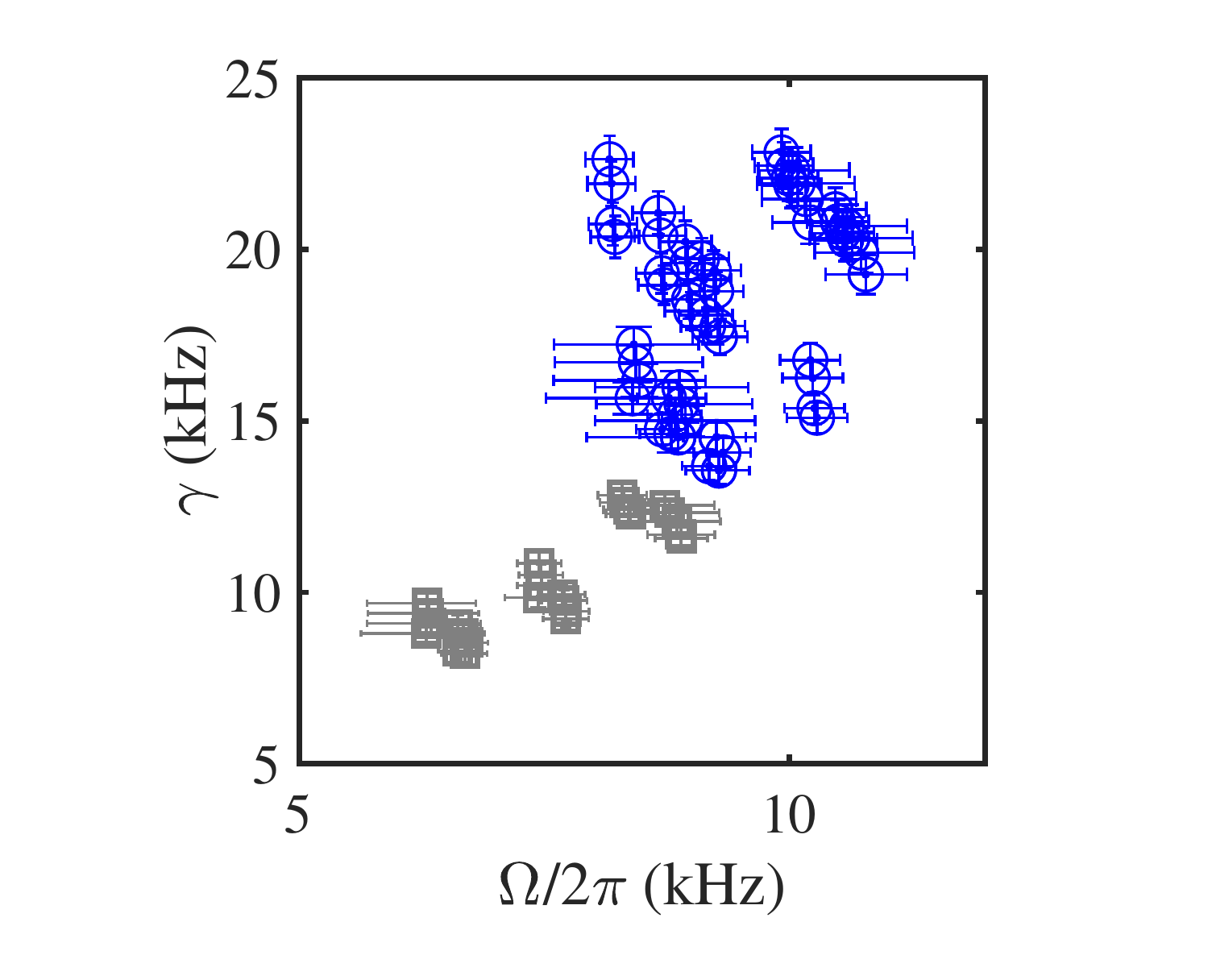}}  	
	\caption{Damping rate ($\gamma$) vs frequency ($\Omega$) obtained in simulations using n, T and $E_r$ (NC) experimental profiles for the set of experimental discharges in table \ref{TabZeff}, 
		and including $C^{+4}$ kinetic impurity. Results for the multi-species kinetic-electron case (blue) and from single-species simulations with adiabatic electrons (gray) are shown (see text for details). The data correspond to  $0.35 < r/a < 0.8$. }
	\label{GammavsOmegaExp}
\end{figure}

In figure \ref{GammavsOmegaExp} we show (in blue) the values of damping rate vs oscillation frequency obtained from the fit of the potential time traces for these simulations. The oscillation frequency is obtained from these simulations and extrapolated to the kinetic-electron case using the formula (\ref{EqFreqiezasfiz}). The damping rates are those directly obtained in the adiabatic-electron multi-species simulations.

In the same figure the results obtained from simulations with single hydrogen ion species and adiabatic electrons (shown in gray color), presented in \cite{Alonso2017}, are included for comparison. Several points corresponding to the values at several radial positions in the range of radial locations of measurements reported in \cite{Alonso2017} ($0.35 < r/a < 0.8$) are plotted in the figure, both for the single-species and the multi-species cases. 

The results in the single-species case, although in qualitative agreement with the experimental measurements, have both damping rates and oscillation frequencies smaller than those estimated from experimental measurements (see  \cite{Alonso2017}). 
%%%%%%%%%%%%%%%%%%%%%%%%%%%%%%%%%%%%%%%%%%%%%%%%%%%%%%%%%%%%%%%%%%%%%%%%%%%%%%%%%%%%%%%
%%%%%%%%%%%%%%%%%%%%%%%%%%%%%%%%%%%%%%%%%%%%%%%%%%%%%%%%%%%%%%%%%%%%%%%%%%%%%%%%%%%%%%%
It is clear from the figure that including an ion impurity in the simulation with realistic impurity concentration values introduces a larger dispersion of results and increases both  the damping rate and the frequency, which makes them closer to the experimental measurements. From this we can conclude that the presence of impurities, even at small concentration, is an important factor to be taken into account for quantitative comparison with experiments.

%%%%%%%%%%%%%%%%%%%%%%%%%%%%%%%%%%%%%%%%%%%%%%%%%%%%%%%%%%%%%%%%%%%%%%%%%%%%%%%%%%%%%%%
%
\section{Discussion and conclusions}\label{SecConcl}
%
%%%%%%%%%%%%%%%%%%%%%%%%%%%%%%%%%%%%%%%%%%%%%%%%%%%%%%%%%%%%%%%%%%%%%%%%%%%%%%%%%%%%%%%
The linear relaxation of a zonal potential perturbation has been studied in the standard configuration of TJ-II as an initial value problem by means of global gyrokinetic simulations with the code EUTERPE. It has been  shown that, in the long-wavelength limit, the oscillation frequency is not dependent on the radial scale of the potential perturbation, as expected from theory.  

We have analyzed the oscillation close to the experimental conditions of the plasmas in which the LFO was detected in TJ-II \cite{Alonso2017}, using experimental density and temperature profiles and the background radial electric field obtained from neoclassical calculations with DKES. It has been shown that including this radial electric field does not modify much the oscillation frequency with respect to the value without electric field. 

The influence of the collisionality has been addressed in simulations with experimental profiles, a single ion species and adiabatic electrons. A pitch-angle collision operator was used in a set of simulations in which the density was changed. The oscillation frequency has been shown to be almost unaffected by collisional processes in this range of collisionality, while the damping rate of the oscillations is largely  affected,  increasing approximately linearly with the density.

The oscillatory relaxation of ZFs in a multi-species plasma has been studied in experimental plasma conditions by means of simulations with two ion species and adiabatic electrons. The frequency has been shown to be reduced when an impurity ion is included at moderate (experimental) concentrations. The oscillation frequency obtained in multi-species simulations in a wide range on impurity concentration has been compared to the values predicted by means of analytical formulas and based on single-species calculations finding a very good agreement.  

It has been shown that the presence of a heavier ion species increases the damping with respect to the case with single-ion species. The contribution of the different inter-species collisions to the damping of the oscillations has been studied in a set of simulations including $C^{+4}$ impurity with specific inter-species collisions deactivated. The collisional damping is found to be much larger than the collisionless one. It was found that the most important contribution to the damping comes from collisions between bulk ions while those between ions with electrons is almost not relevant. The collisions of bulk ions with C impurity ions are important for small values of the effective charge while the contribution of C-C ions increases for effective charge above 1.7.
 
The oscillation frequency and damping rate has been studied in experimental plasma conditions in which the LFO was detected in TJ-II in simulations with adiabatic electrons and including the dominant impurity ion in these TJ-II plasmas, $C^{+4}$, with concentrations in the range of estimations of $Z_{eff}$ from SXR experimental measurements for those plasma discharges. The resulting oscillation frequencies and damping rates are much closer to the experimental measurements than previous estimations based on single-species simulations.

Two important conclusions can be extracted from this work. The first conclusion is that the ZF oscillation frequencies and damping rates obtained in simulations including multi-species plasmas with realistic concentration of impurities is in quantitative agreement with the experimental measurements in TJ-II \cite{Alonso2017}. It has to be taken into account that these results were obtained in multi-species simulations with adiabatic electrons. A small contribution to the damping rate can be expected from kinetic electrons. However, the contribution of electrons to the oscillation frequency could be reduced if they are in a very collisional regime.

A second important, and more general,  conclusion that can be drawn is that the ZF oscillation frequency in a multi-species plasma can be accurately estimated from single-species calculation/simulations, which can be carried out at a much smaller computational cost. 

In ref.~\cite{Monreal2017} analytical formulas were derived for the oscillation frequency in general stellarator geometry that can be evaluated at a cheaper cost than gyrokinetic simulations. However, in this work we have used simulations rather than semi-analytical calculations, for all the calculations. The reason is that the expressions derived in \cite{Monreal2017} do not take into account the radial electric field, nor the gradients in the density and temperature profiles and collisional processes, which all play a role in the quantitative comparison with experimental measurements. It should be noted that the analytical expressions still have interest because they can capture the influence of the magnetic configuration at a cheaper computational  cost than the gyrokinetic simulations. In addition, we have shown in this work that the oscillation frequency is not much modified by the background electric field nor the collisional processes under the experimental conditions described in \cite{Alonso2017}, however this is  can not be assumed true in general. 

 As can be clearly seen in expression (\ref{EqLowFrequencyDef}) derived in \cite{Monreal2017} the ZF oscillation frequency involves quantities that are averaged over the full flux surface. This indicates that the minimum computational domain to study this problem in a stellarator is a full-flux-surface, yet radially-local, physical domain. Comparison of ZF relaxation in different computational domains are in progress and have already evidenced important discrepancies between flux tube (radially local) and radially global calculations \cite{Smoniewski2018}. Thinking of non-linear turbulence simulations in which the zonal flow response can be relevant for the turbulence saturation the computational domain used in the simulation can also make an important difference.

This low frequency ZF oscillation is expected to be particularly interesting in W7-X configurations, because according to calculations it has a large amplitude and small collisionless damping rate \cite{Monreal2017} and could affect the turbulent transport level \cite{XanthopoulosPRL2011}. As  W7-X operates in a less collisional regime than TJ-II the damping will be much smaller in that case, and then the oscillation should be easier to detect than in TJ-II. Estimations of effective charge in OP1.1 experimental campaign give $1.5 < Z_{eff} < 5.5$. Then, in order to make a quantitative comparison between calculations/simulations and experimental measurements the corrections due to the presence of multiple ion species and kinetic electrons here studied will be required.

\section{Acknowledgements}
This work has been partially funded by the Ministerio de Econom\'ia y Competitividad of Spain under project ENE2015-70142-P. The
authors thankfully acknowledge the computer resources, technical expertise and assistance provided by the Barcelona Supercomputing
Center-Centro Nacional de Supercomputaci\'on and the CIEMAT computing center.

This work has been carried out within the framework of the EUROfusion Consortium and has received funding from the Euratom research and training programme 2014-2018 under grant agreement No 633053. The views and opinions expressed herein do not necessarily reflect those of the European Commission.

%%%%%%%%%%%%%%%%%%%%%%%%%%%%%%%%%%%%%%%%%%%%%%%%%%%%%%%%%%%%%%%%%%%%%%%%%%%%%%%%%%%%%%%
%
%\section{References}
%
%%%%%%%%%%%%%%%%%%%%%%%%%%%%%%%%%%%%%%%%%%%%%%%%%%%%%%%%%%%%%%%%%%%%%%%%%%%%%%%%%%%%%%%%
%\bibliographystyle{unsrt}%,alpha}
%\bibliography{Sanchez_ISHW2017_2_PPCF_v1.1.bib}{}
%\thebibliography

%\printbibliography

%%%%%%%%%%%%%%%%%%%%%%%%%%%%%%%%%%%%%%%%%%%%%%%%%%%%%%
\section*{References}
%%%%%%%%%%%%%%%%%%%%%%%%%%%%%%%%%%%%%%%%%%%%%%%%%%%%%%

\end{document}